\documentclass[aps,prl,twocolumn,amsmath,amssymb,nofootinbib,superscriptaddress]{revtex4}
\usepackage{color}
\usepackage{soul}
\usepackage{amsfonts}
\usepackage{amsmath}
\usepackage{amssymb}
\usepackage{mathrsfs}
\usepackage{graphicx}%
\usepackage{physics}
\usepackage{multirow}

\newcommand{\tim}[1]{{\color{black}{#1}}}

\providecommand{\U}[1]{\protect\rule{.1in}{.1in}}

\begin{document}
\title{\tim{Emulating quantum teleportation of a Majorana zero mode qubit}}

\author{He-Liang Huang}
\affiliation{Hefei National Laboratory for Physical Sciences at the Microscale and Department of Modern Physics, University of Science and Technology of China, Hefei 230026, China}
\affiliation{Shanghai Branch, CAS Center for Excellence in Quantum Information and Quantum Physics, University of Science and Technology of China, Shanghai 201315, China}
\affiliation{Shanghai Research Center for Quantum Sciences, Shanghai 201315, China}

\author{Marek Naro$\dot{\text{z}}$niak}
\affiliation{New York University Shanghai, 1555 Century Ave, Pudong, Shanghai 200122, China}
\affiliation{Department of Physics, New York University, New York, NY 10003, USA}

\author{Futian Liang}
\affiliation{Hefei National Laboratory for Physical Sciences at the Microscale and Department of Modern Physics, University of Science and Technology of China, Hefei 230026, China}
\affiliation{Shanghai Branch, CAS Center for Excellence in Quantum Information and Quantum Physics, University of Science and Technology of China, Shanghai 201315, China}
\affiliation{Shanghai Research Center for Quantum Sciences, Shanghai 201315, China}

\author{Youwei Zhao}
\author{Anthony D. Castellano}
\author{Ming Gong}
\author{Yulin Wu}
\author{Shiyu Wang}
\author{Jin Lin}
\author{Yu Xu}
\author{Hui Deng}
\author{Hao Rong}
\affiliation{Hefei National Laboratory for Physical Sciences at the Microscale and Department of Modern Physics, University of Science and Technology of China, Hefei 230026, China}
\affiliation{Shanghai Branch, CAS Center for Excellence in Quantum Information and Quantum Physics, University of Science and Technology of China, Shanghai 201315, China}
\affiliation{Shanghai Research Center for Quantum Sciences, Shanghai 201315, China}

\author{Jonathan P. Dowling \footnote{In memory of Jonathan P. Dowling}}
\affiliation{Hearne Institute for Theoretical Physics, Department of Physics and Astronomy, Louisiana State University, Baton Rouge, Louisiana 70803, USA}
\affiliation{Hefei National Laboratory for Physical Sciences at Microscale and Department of Modern Physics,\\
University of Science and Technology of China, Hefei, Anhui 230026, China}
\affiliation{NYU-ECNU Institute of Physics at NYU Shanghai, 3663 Zhongshan Road North, Shanghai 200062, China}

\author{Cheng-Zhi Peng}

\affiliation{Hefei National Laboratory for Physical Sciences at the Microscale and Department of Modern Physics, University of Science and Technology of China, Hefei 230026, China}
\affiliation{Shanghai Branch, CAS Center for Excellence in Quantum Information and Quantum Physics, University of Science and Technology of China, Shanghai 201315, China}
\affiliation{Shanghai Research Center for Quantum Sciences, Shanghai 201315, China}

\author{Tim Byrnes}
\affiliation{New York University Shanghai, 1555 Century Ave, Pudong, Shanghai 200122, China}
\affiliation{NYU-ECNU Institute of Physics at NYU Shanghai, 3663 Zhongshan Road North, Shanghai 200062, China}
\affiliation{State Key Laboratory of Precision Spectroscopy, School of Physical and Material Sciences,East China Normal University, Shanghai 200062, China}
\affiliation{Department of Physics, New York University, New York, NY 10003, USA}

\author{Xiaobo Zhu}
\email{xbzhu16@ustc.edu.cn}
\author{Jian-Wei Pan}
\affiliation{Hefei National Laboratory for Physical Sciences at the Microscale and Department of Modern Physics, University of Science and Technology of China, Hefei 230026, China}
\affiliation{Shanghai Branch, CAS Center for Excellence in Quantum Information and Quantum Physics, University of Science and Technology of China, Shanghai 201315, China}
\affiliation{Shanghai Research Center for Quantum Sciences, Shanghai 201315, China}

\begin{abstract}
Topological quantum computation based on anyons is a promising approach to achieve fault-tolerant quantum computing. The Majorana zero modes in the Kitaev chain are an example of non-Abelian anyons where braiding operations can be used to perform quantum gates. \tim{Here we perform a quantum simulation of topological quantum computing, by teleporting a qubit encoded in the Majorana zero modes of a Kitaev chain. The quantum simulation is performed by mapping the Kitaev chain to its equivalent spin version, and realizing the ground states in a superconducting quantum processor. The teleportation transfers the quantum state encoded in the spin-mapped version of the Majorana zero mode states between two Kitaev chains. The teleportation circuit is realized using only braiding operations,
and can be achieved despite being restricted to Clifford gates for the Ising anyons.} The Majorana encoding is a quantum error detecting code for phase flip errors, which is used to improve the average fidelity of the teleportation for six distinct states from $70.76 \pm 0.35 \% $ to $84.60 \pm 0.11 \%$, well beyond the classical bound in either case.
\end{abstract}

\date{\today}

\setcounter{figure}{0}
\setcounter{equation}{0}
\makeatletter

\newcommand{\manuallabel}[2]{\def\@currentlabel{#2}\label{#1}}
\makeatother
\manuallabel{fig1}{1}
\manuallabel{fig2}{2}
\manuallabel{fig3}{3}

\maketitle

\tim{One of the most attractive ways of performing fault-tolerant quantum computing \cite{shor1996fault,gottesman1998theory,steane1999efficient,aharonov1999fault,preskill1998fault,gottesman2010introduction,devitt2013quantum,campbell2017roads} is topological quantum computing \cite{kitaev2003fault,freedman2003topological,preskill2004topological,nayak2008non,pachos2012introduction,sarma2015majorana,lahtinen2017short}.  In topological quantum computing, the quantum information is stored in the states of anyons, which have a non-trivial effect on the total state when they are interchanged. The logical states of the anyons form a subspace distinguishing the error-free space to those with errors, and errors are suppressed via the topological gap. For non-Abelian anyons, their braiding can be used to construct elementary quantum gates for quantum computing. The resulting quantum gate is only dependent upon the topology of the braiding path, thus small imperfections in the braiding can be tolerated as long as the operation is topologically equivalent.}

One example of a non-Abelian anyon is the Majorana zero mode (MZM) \cite{kitaev2001unpaired,alicea2012new,leijnse2012introduction,beenakker2013search,sarma2015majorana,lutchyn2018majorana}.  MZMs are zero energy excitations that occur typically in low-dimensional topological superconductors. Two physical systems where MZMs have been intensely investigated are fractional quantum Hall systems \cite{willett1987observation,sarma2005topologically,willett2009measurement,willett2013magnetic} and semiconductor nanowires \cite{mourik2012signatures,lutchyn2010majorana,oreg2010helical}.  \tim{To date, many experiments have been conducted to find the evidence for the existence of Majorana fermions, however, the key feature of topological protection has not yet been demonstrated \cite{sola2020majorana}. An elementary model that possesses MZMs is the Kitaev chain consisting of $ N $ fermions with Hamiltonian \cite{kitaev2001unpaired}
\begin{align}
H =  \sum_{n=1}^{N-1}  \Delta (c_{n} c_{n+1} + c_{n+1}^\dagger  c_{n}^\dagger)  -t ( c_{n+1}^\dagger c_{n}  + c_{n}^\dagger c_{n+1}) - \mu c_{n}^\dagger c_{n} ,
\label{kitaevham1}
\end{align}
where $  c_{n} $ is a fermionic annihilation operator on site $ n $, and $ t $ is the hopping energy, $ \Delta $ is the superconducting gap, and $ \mu $ is a chemical potential.
For finite $ N $ and working in the limit $ \Delta = t $ and $ \mu = 0 $,} the model has a degenerate ground state,  corresponding to the presence or absence of a pair of MZMs, and can be used to encode the state of a qubit.  By braiding one of the MZMs with another, quantum gates on the encoded quantum information may be performed, thereby forming the basis for topological quantum computing.

\tim{While the direct realization of topological quantum computing based on MZMs is still out of reach of present day technology, the study of the Majorana physics is now possible by way of quantum simulation in other controllable systems such as superconducting and ion-trap systems \cite{huang2020superconducting}.  Mapping the Kitaev chain to a spin model via the Jordan–Wigner transformation, the Hamiltonian (\ref{kitaevham1}) takes the form of a one-dimensional transverse field Ising model, which has made it attractive to numerous proposals for simulating its equivalent dynamics.
Xu, Pachos, and Guo implemented the spin version of MZMs states in a Kitaev chain, and braiding of the effective anyons was demonstrated to realize one qubit gates with imaginary time evolution \cite{xu2016simulating,xu2018photonic}. Several works have also demonstrated the path-independent nature of braiding anyonic excitations in the toric code \cite{lu2009demonstrating,pachos2009revealing,zhong2016emulating,dai2017four,song2018demonstration,liu2019demonstration}. Simulating the physics of Majorana fermions by artificially constructed lattice models can provide additional insight into the nature of the quantum states, such as allowing one to tap into the existing pool of ideas on Majorana-based quantum computation \cite{tserkovnyak2011universal, mezzacapo2013topological}. To date, such studies have been restricted to examining the basic properties of anyons, and we are not aware of any quantum simulation of topological quantum computing involving more than one encoded qubit.}

\begin{figure}
\begin{center}
\includegraphics[width=\linewidth]{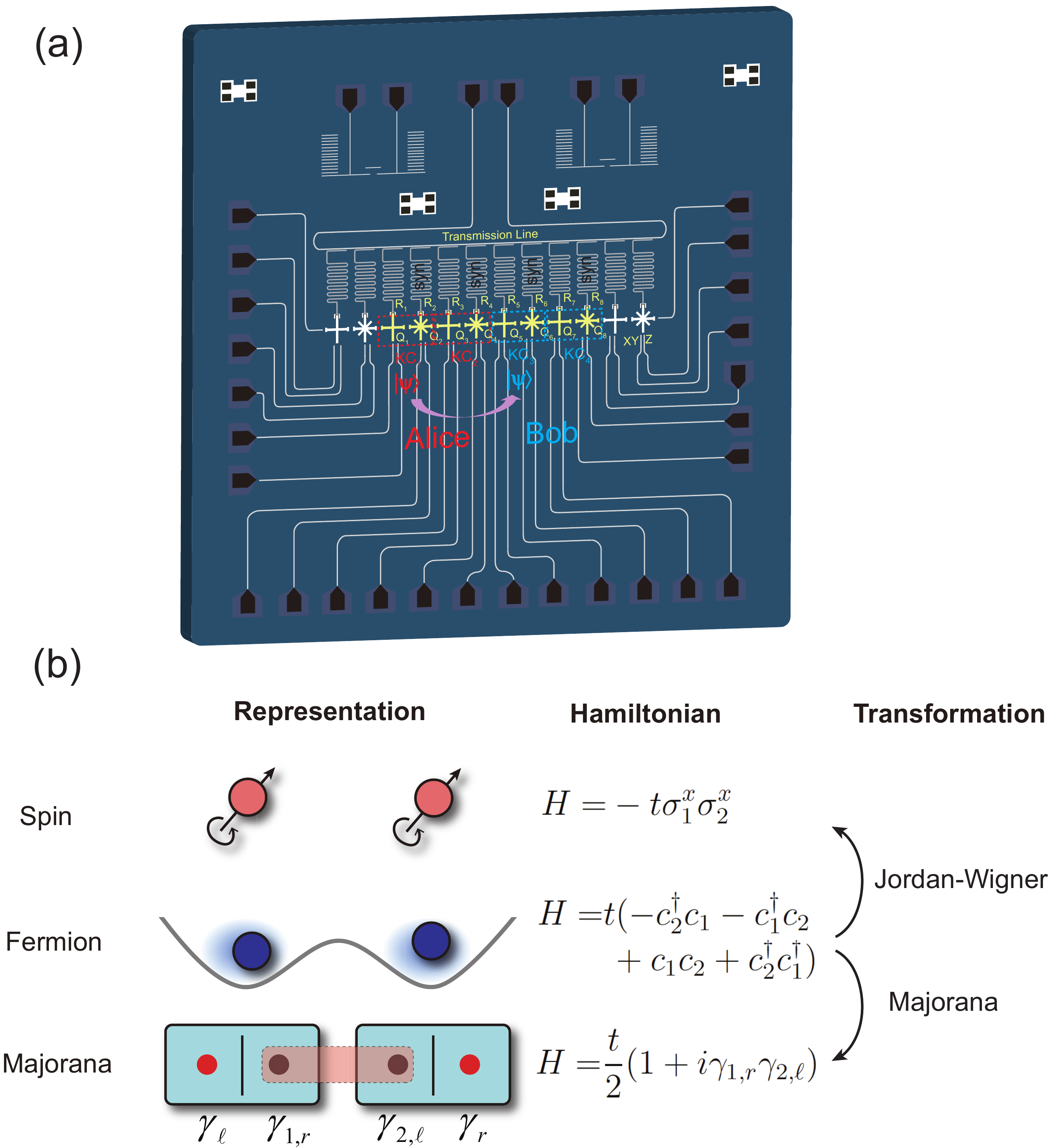}
\end{center}
\caption{\linespread{1}{Experimental configuration and \tim{the mapping of Kitaev chain to its spin and Majorana representations.}
(a) The superconducting quantum processor. We choose eight adjacent qubits labelled with $Q_1$ to $Q_8$ from the 12-qubit processor to perform the experiment. Qubits $Q_1$ to $Q_4$ and $Q_5$ to $Q_8$ are held by Alice and Bob, respectively. Pairs of qubits form a \tim{spin-mapped} Kitaev chain ($KC$), each which encode a single logical qubit. Each qubit couples to a resonator for state readout, marked by $R_1$ to $R_8$. After decoding, the resonators marked by ``syn'' are syndrome measurements to detect phase flip errors in the qubits. An encoded qubit is teleported from $KC_1 $ to $ KC_3 $. (b) Mapping between spin, fermions, and Majorana modes. The pairing of Majorana modes in the topologically non-trivial regime are indicated by the dotted ovals.  In the topologically non-trivial phase, the MZMs are present at the ends of the chain.} 
\label{fig1}}
\end{figure}

\tim{In this paper, we investigate the feasibility of topological quantum computing by simulating the quantum teleportation \cite{bennett1993teleporting} of a MZM state of the Kitaev chain on superconducting qubits.}
We realize four spin-mapped Kitaev chains using eight superconducting qubits (see Fig. \ref{fig1}).  Each chain, consisting of two physical qubits, encodes a single logical qubit, corresponding to the \tim{spin-mapped} MZM states. In the teleportation, Alice is in possession of two of the Kitaev chains, and Bob holds the two other chains.  The teleportation then transfers a single logical qubit, encoded as the \tim{spin-mapped} MZM states, from Alice to Bob. One of the well-known issues of quantum computing based on MZMs is that braiding operations only allow for a discrete number of Clifford gates, which is insufficient for universal quantum computation \cite{freedman2002modular,preskill2004topological,stern2013topological}.  Fortunately in the teleportation protocol, only Clifford gates are required, such that it can be completed entirely with braiding operations. \tim{In addition to demonstrating the feasibility of anyonic quantum computing, we also show the error protection capabilities of MZMs. The protection of quantum information via error correcting codes has been demonstrated in many past works \cite{yao2012experimental,nigg2014quantum,kelly2015state,corcoles2015demonstration,ofek2016extending,linke2017fault,takita2017experimental,rosenblum2018fault,harper2019fault, andersen2020repeated}, including those based on topological states \cite{nigg2014quantum,andersen2020repeated}. We show that the spin-mapped MZM states are capable of detecting phase errors, allowing us to improve the teleportation fidelity significantly.}


We first give a brief review of anyonic quantum computing with MZMs.
Each fermion is written in terms of two Majoranas according to the definition
\begin{align}
\gamma_{n,\ell}& = c_n + c_n^\dagger \nonumber \\
\gamma_{n,r} & = -i c_n + i c_n^\dagger
\end{align}
where $ n $ is an integer labeling the fermions, and the $ \ell, r $ label the two types of Majoranas, which correspond to the real and imaginary part of the fermion operator, denoted by the left and right boxes in Fig. \ref{fig1}(b) respectively.  Let us denote  $ | 0_L \rangle $  a ground state of the Hamiltonian (\ref{kitaevham1}), taken as the state with no Majorana modes throughout the chain.  The nature of the Kitaev Hamiltonian is such that applying the fermion creation operator
\begin{align}
f^\dagger  = \frac{1}{2} ( \gamma_{1,\ell} - i \gamma_{N,r}) ,
\end{align}
consisting of two Majorana edge modes at the ends of the lattice, produces another orthogonal degenerate state.  These two states $| 0_L \rangle$ and $ |1_L \rangle \equiv f^\dagger |0 \rangle $ are the MZM states and are used as the logical qubit states. A minimal implementation of the Kitaev chain Hamiltonian (\ref{kitaevham1}) consists of two fermions $ N = 2$.  Under the Jordan-Wigner mapping, the Hamiltonian takes a form $ H = - t \sigma^x_1 \sigma^x_2 $ \tim{for $ \Delta = t, \mu = 0 $,} and the two MZM states are
\begin{align}
| 0_L \rangle & = \frac{1}{\sqrt{2}} ( \ket{++}  + \ket{--} ) \nonumber \\
| 1_L \rangle & = \frac{1}{\sqrt{2}} ( \ket{++} - \ket{--} ) .
\label{logicalmzm1}
\end{align}

To encode $ M $ logical qubits, one then prepares $ M $ Kitaev chains, each with the Hamiltonian (\ref{kitaevham1}). Let us label the MZMs from the $ m $th chain as
\begin{align}
\gamma_{\ell}^{(m)} & \equiv \gamma_{1,\ell}^{(m)} \nonumber \\
\gamma_{r}^{(m)} & \equiv \gamma_{N,r}^{(m)},
\end{align}
such that we only label the left-most and right-most Majorana mode in the chain, which are the MZMs.  An MZM, on the $m$th chain that is in the left- or right-most position $ \sigma \in \{ \ell, r \} $, can be braided with another labeled by $ (m', \sigma') $.  The effect of this is to apply the unitary braid operator \cite{nayak19962n,ivanov2001non}, defined as
\begin{align}
B_{(m, \sigma) (m', \sigma')} = e^{\pi \gamma_{\sigma}^{(m)} \gamma_{\sigma'}^{(m')} /4} = \frac{1}{\sqrt{2}} ( 1 + \gamma_{\sigma}^{(m)} \gamma_{\sigma'}^{(m')} ) .
\label{braidingop}
\end{align}
For two logical qubits, there are four MZMs, and therefore there are $  {4 \choose 2 } = 6 $ possible braiding operations 
(see Fig.~S4 in Supplemental Material). Due to the non-Abelian nature of MZMs, these produce gate operations on MZM states. In our circuit, we utilize the fact that braiding MZMs on the same chain produce a logical $ \sqrt{Z} $ operation, and braiding adjacent MZMs on two different chains produce a logical $ \sqrt{X_1 X_2} $.

\begin{figure*}
\begin{center}
\includegraphics[width=0.9\linewidth]{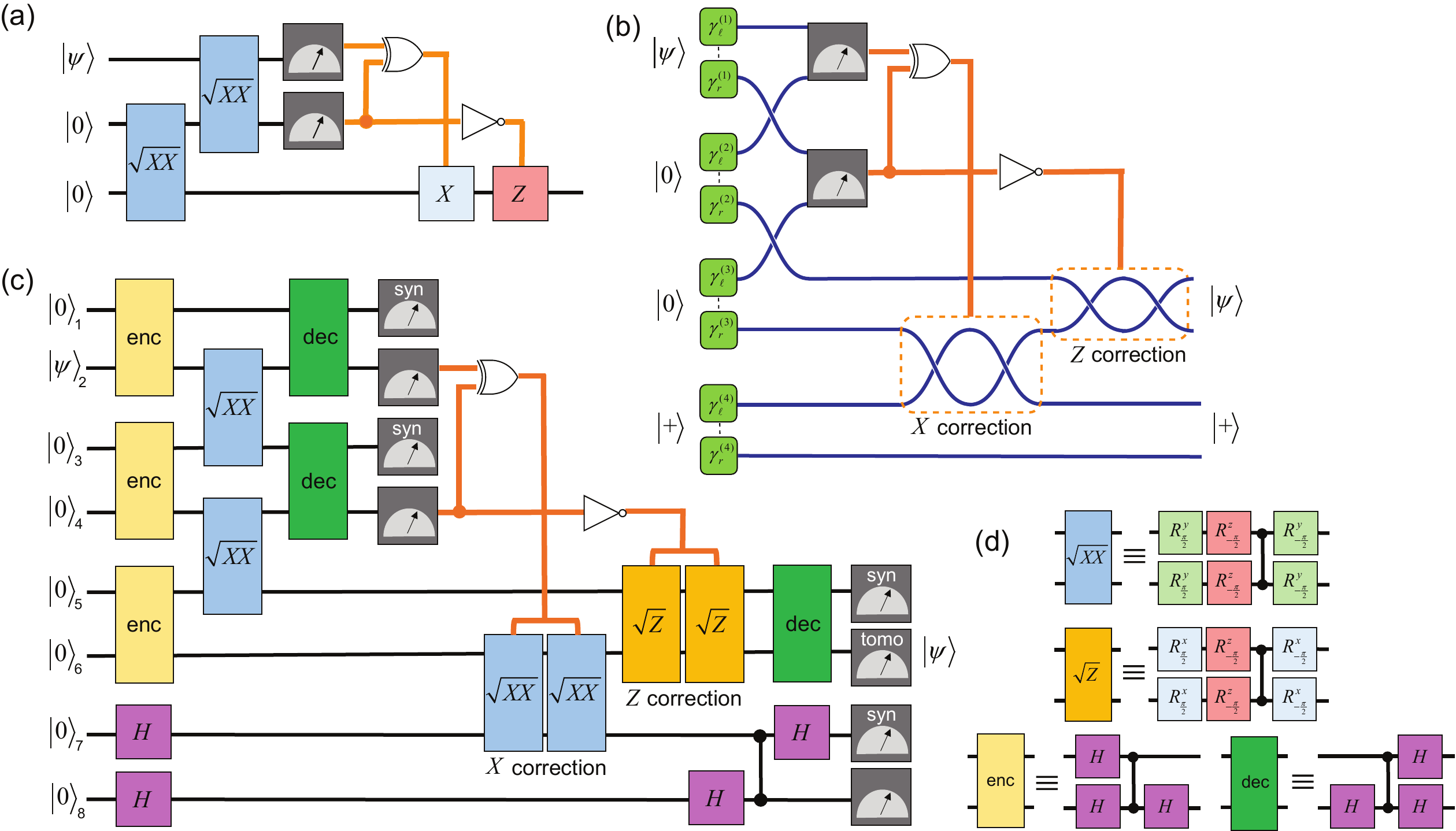}
\end{center}
\caption{\tim{Quantum circuits for simulating the teleportation of a MZM encoded qubit}.  (a) The modified quantum teleportation scheme. (b) The braiding sequence for MZMs that performs the quantum circuit in (a).  (c) The corresponding \tim{spin-mapped} qubit circuit of the MZM braiding sequence shown in (b).   All measurements are performed in the $ |0 \rangle, | 1\rangle $, which the exception of the measurement on qubit 6, where tomography (``tomo'') is performed.  The measurements marked with ``syn'' are syndrome measurements, where single qubit phase errors are detected for a measurement outcome of $ | 1 \rangle $.  (d) The gate decompositions for the braiding, encoding, and decoding gates in (c). 
In all figures, $ | \psi \rangle $ is the state to be teleported.  Black lines connecting the quantum gates denote qubits, dark blue lines denote MZMs, and orange lines denote classical information transfer.
\label{fig2}}
\end{figure*}

The standard quantum teleportation circuit consists of a sequence of Hadamard and CNOT gates \cite{nielsen2002quantum}, which are not directly available by braiding operations. To match the gates that are available with braiding operations as closely as possible, we design a modified teleportation scheme (see Fig.~\ref{fig2}(a) and Supplemental Material). The protocol proceeds in a similar way to the standard teleportation circuit, except that the classically transmitted quantum correction (``classical correction'' for short) is done according to the modified rules shown in the classical circuit of Fig.~\ref{fig2}(a). Using this modified teleportation circuit, the equivalent version with MZMs can be constructed entirely using the six available braiding gates. The one gate that is not directly available as a braiding gate in the circuit of Fig.~\ref{fig2}(a)  is the $ X $-gate for classical correction. No combination of the six braiding operations in Fig.~S4 can produce a single qubit $ X $-gate.  However, by adding an extra ancilla MZM qubit ($m=4$) prepared in the eigenstate with $ X_4 = +1 $, and applying the braiding operation for the logical  $ \sqrt{X_3 X_4} $ twice, we can perform an $X_3 $ gate.  In this way all gates appearing in the teleportation circuit can be performed natively using only braiding operations (Fig.~\ref{fig2}(b)).
Using a minimal implementation of the Kitaev chain with $ N = 2 $ fermions, and performing a Jordan-Wigner transformation, we convert the MZM teleportation circuit (Fig.~\ref{fig2}(b)) into the equivalent 8 qubit version as shown in Fig.~\ref{fig2}(c).

In addition to the braiding operations,
we require encoding and decoding operations to prepare the logical \tim{spin-mapped} MZM qubit states of (\ref{logicalmzm1}).  The encoder takes an arbitrary qubit state and an auxiliary qubit in the state $ | 0 \rangle $ and produces its associated logical \tim{spin-mapped} MZM qubit state
\begin{align}
U_{\text{enc}} | 0 \rangle (\alpha |0 \rangle + \beta |1 \rangle ) = \alpha |0_L \rangle + \beta  |1_L \rangle ,
\end{align}
which can be performed using elementary gates and the definitions (\ref{logicalmzm1}). Here $ \alpha, \beta $ are arbitrary complex coefficients such that $ |\alpha |^2 + | \beta |^2 = 1 $. The gate decompositions for the braiding gates, encoder, and decoder $ U_{\text{dec}} = U_{\text{enc}}^\dagger $ are shown in Fig.~\ref{fig2}(d).

We choose eight adjacent qubits from a 12-qubit superconducting quantum processor \cite{gong2019genuine,huang2020experimental} to implement the quantum circuit of Fig.~\ref{fig2}(c).
The average fidelities of single-qubit gates and the controlled-$Z$ gate are approximately 0.9994 and $0.985$, respectively. The six input states of $ |0 \rangle, |1 \rangle, |+ \rangle,  |- \rangle, |+i  \rangle, |-i  \rangle $, corresponding to pairs of eigenstates of the Pauli $ \sigma^z, \sigma^x, \sigma^y $ operators are prepared on qubit 2 as the input state for teleportation. To perform the classical correction steps, we run four versions of the circuit with and without each of the $ X $ and $ Z $ classical correction gates.  Then given a particular measurement outcome on qubits 2 and 4, the correct circuit for that outcome is selected.
To perform the tomography measurement of the teleported state on qubit 6, we repeat the circuit by making measurements in the $ X, Y, Z $ basis such that the state can be tomographically reconstructed. Each of the circuit variants were run a total of 40000 times for statistics.

\begin{figure}
\begin{center}
\includegraphics[width=\linewidth]{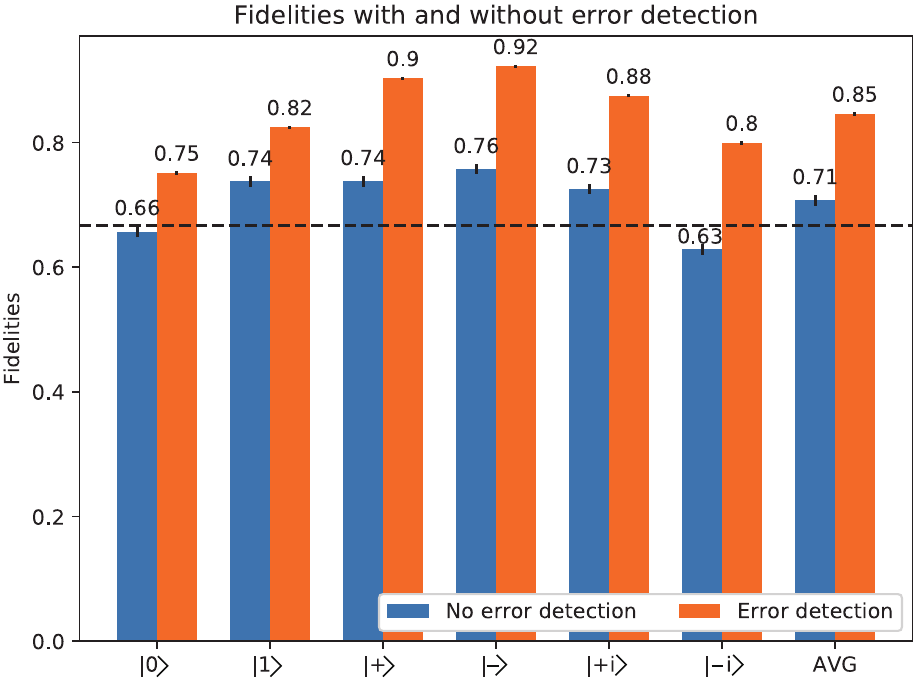}
\end{center}
\caption{Teleportation fidelities with and without error syndrome detection. The fidelity is calcuated according to $ F = \langle \psi | \rho | \psi \rangle $, where $ | \psi \rangle = \{ |0 \rangle, | 1 \rangle, | + \rangle, | - \rangle, | +i \rangle , | -i \rangle  \} $ are the ideal states to be teleported.  The dashed line denotes the $ F = 2/3$ threshold. The error bars denote one standard deviation.
\label{fig3}}
\end{figure}

Figure~\ref{fig3} shows the teleportation fidelities for each of the six input states.   First we average over all measurement outcomes on qubits $1,3,5,7,8$, which corresponds to ignoring all error syndrome measurements and any changes in the ancilla MZM qubit. We find the average fidelity of the six states is $70.76 \pm 0.35 \% $, which is above the classical limit of 2/3
\cite{massar2005optimal} by 11 standard deviations. We have performed an explicit simulation of the circuit shown in Fig.~\ref{fig2}(c) including dephasing and gate errors, and obtain good agreement between the experimentally obtained errors (see Supplemental Material). We note that the experiment further suffers from readout errors, which are expected to further degrade the theoretical fidelities.
From the operations on qubit 7 and 8 it is apparent that the final state should be in the state $ | 00 \rangle $, which is consistent with the fact that the role of these qubits are only to be in the $ X = 1 $ eigenstate. We experimentally obtain the probability of getting the $ | 00 \rangle $ state is $97.98\%$, consistent with this expectation.

\tim{One of the benefits of encoding quantum information with MZMs is that it allows for a natural way of protecting against errors.  As stated in Kiteav's original paper \cite{kitaev2001unpaired} introducing the model (\ref{kitaevham1}), the MZM encoding is resilient against phase-flip errors because of they correspond to non-local fermion interactions, and bit-flip errors because they corresponds to a parity non-conserving process, which are both unlikely to occur naturally.  Under the spin-mapping, the protection against bit-flip errors is lost, but protection against phase-flip errors is still present (see Supplemental Material). This can be easily seen by examining the logical states after a phase-flip
\begin{align}
 |\tilde{0}_L \rangle & = \sigma^z_1  | 0_L \rangle =  \sigma^z_2 | 0_L \rangle =
\frac{1}{\sqrt{2}} ( \ket{-+} + \ket{+-} ) \nonumber \\
 |\tilde{1}_L \rangle & = \sigma^z_1  | 1_L \rangle =  - \sigma^z_2 | 1_L \rangle =
\frac{1}{\sqrt{2}} ( \ket{-+} - \ket{+-} ) .
\end{align}
The states $  |\tilde{0}_L \rangle,  |\tilde{1}_L \rangle $ span an orthogonal subspace to that spanned by the logical states, and are produced when any single qubit phase error occurs. Using the relation
\begin{align}
U_{\text{dec}} (\alpha |\tilde{0}_L \rangle + \beta  |\tilde{1}_L \rangle )= | 1 \rangle (\alpha |0 \rangle + \beta |1 \rangle )  ,
\label{errorenc}
\end{align}
we can see that the measurements outcomes of $ | 1 \rangle $ on the auxiliary qubits allows one to deduce that a phase flip error has occurred on any of the qubits.  This constitutes an error detecting code \cite{vaidman1996error,grassl1997codes,preskill1998fault,jordan2006error}, which can be used to passively improve the fidelity of the circuit by discarding any results where errors have occurred
\cite{kelly2015state,corcoles2015demonstration,linke2017fault,takita2017experimental,rosenblum2018fault,harper2019fault}. }



\begin{figure}
\begin{center}
\includegraphics[width=\linewidth]{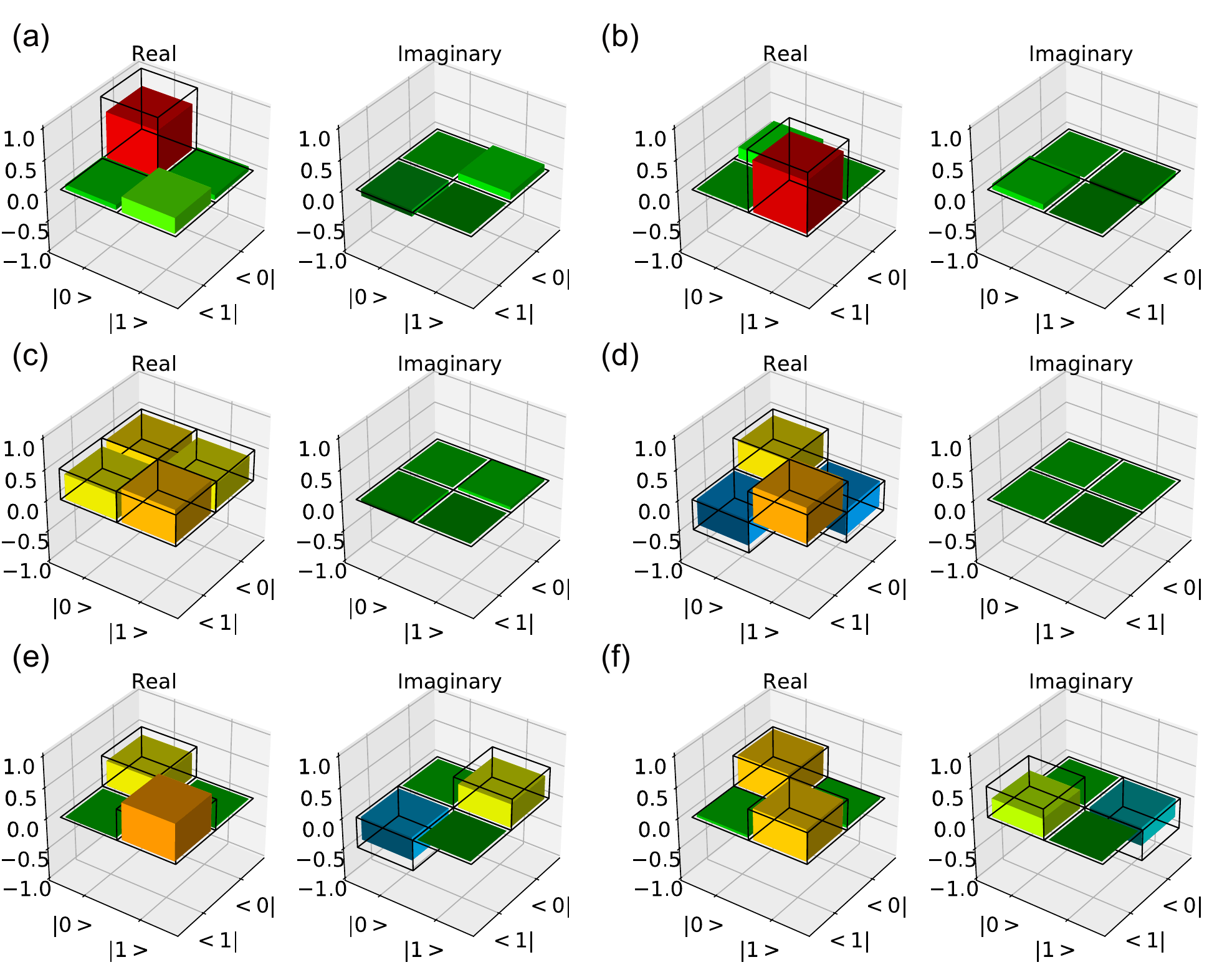}
\end{center}
\caption{Tomography of the final teleported state after using the error syndrome measurements. The initial state prepared on qubit 2 is (a) $ | 0 \rangle $,  (b) $ | 1 \rangle $,  (c) $ | + \rangle $,  (d) $ | - \rangle $,  (e) $ | +i \rangle $,  (f) $ | -i \rangle $. Frames show ideal teleportation states, colored bars shows the experimentally determined state.
\label{fig4}}
\end{figure}

Figure \ref{fig3} shows the teleportation fidelities using the error syndrome measurements on qubits $1,3,5,7$.  A measurement of the state $ |1 \rangle $ on any of these qubits signals that at least one phase flip error has occurred, such that it is removed from the data set.  All results on the ancilla qubit 8 are included. We observe that the fidelities of the telportation improve significantly for all states, with an average of $84.60 \pm 0.11 \%$ for the six states.  This further increases the average fidelity beyond the classical bound by over 163 standard deviations.
The final teleported state for the error syndrome improved state is tomographically reconstructed using a maximum likelihood estimator of the density matrix and shown in Fig. \ref{fig4}. We see that the states of the six input states are very well-reproduced, demonstrating that the teleportation is being performed correctly over the six mutually unbiased basis states.

In summary, \tim{we have performed a quantum simulation of the teleportation of a qubit encoded as the MZM states of the Kitaev chain.} The teleportation circuit is performed entirely using braiding operations of the MZMs, including the quantum gates for classical correction.  In our teleportation circuit we were careful to be faithful to the braiding process of the MZMs in the sense that no gate simplifications were performed in the circuit Fig. \ref{fig2}(c).  \tim{This demonstrates via an equivalent spin encoding that
a non-trivial quantum circuit can be performed using a topological quantum computing sequence using braids. In addition, numerous demonstrations of teleportation have been performed to date in qubit \cite{bouwmeester1997experimental,boschi1998experimental,riebe2004deterministic,barrett2004deterministic,ma2012quantum,baur2012benchmarking,pfaff2014unconditional,ren2017ground} and higher dimensional systems \cite{furusawa1998unconditional,zhang2006experimental,krauter2013deterministic,wang2015quantum,luo2019quantum}, but never in combination with quantum error correction. Purely from the perspective of the spin formulation, our work can also be viewed as a demonstration of an encoded qubit teleportation.}


\tim{The benefit of the MZM encoding in the original fermion model is protection against both bit- and phase-flip errors \cite{kitaev2001unpaired}.  By performing the spin-mapping, the protection against bit-flip errors is lost, as evident by examining the states (\ref{logicalmzm1}).  However, the protection against phase-flip errors is still present, and we explicitly demonstrated the enhancement in fidelity of the teleportation by postselecting states without detected errors. Thus, our results also indicate the feasibility of phase error protection in MZM-based topological quantum computing. Longer chains are expected to enhance the phase error protection, but will make the states more susceptible to bit-flip errors due to a higher probability of an error occurring somewhere on the chain.  Thus while we do not expect that the spin encoding benefits from larger $ N $, in the original fermion model we expect that the logical states should have an enhanced protection.   In addition to the error detection performed here, with the addition of a topological gap to energetically separate the logical space from the error space, errors could be actively suppressed, further improving the error protection. }

\begin{acknowledgments}
The authors thank the Laboratory of Microfabrication, University of Science and Technology of China, Institute of Physics CAS, and National Center for Nanoscience and Technology for supporting the sample fabrication. The authors also thank QuantumCTek Co., Ltd., for supporting the fabrication and the maintenance of room-temperature electronics. \textbf{Funding}: This research was supported by the National Key Research and Development Program of China (Grants No. 2017YFA0304300), NSFC (Grants No. 11574380, No. 11905217), the Chinese Academy of Science and its Strategic Priority Research Program (Grants No. XDB28000000), the Science and Technology Committee of Shanghai Municipality, Shanghai Municipal Science and Technology Major Project (Grant No.2019SHZDZX01), and Anhui Initiative in Quantum Information Technologies. T.B. is supported by the National Natural Science Foundation of China (61571301,D1210036A); the NSFC Research Fund for International Young Scientists (11650110425, 11850410426); NYU-ECNU Institute of Physics at NYU Shanghai; the Science and Technology Commission of Shanghai Municipality (17ZR1443600, 19XD1423000); the China Science and Technology Exchange Center (NGA-16-001); and the NSFC-RFBR Collaborative grant (81811530112). H.-L. H. is supported by the Open Research Fund from State Key Laboratory of High Performance Computing of China (Grant No. 201901-01), NSFC (Grants No. 11905294), and China Postdoctoral Science Foundation.

H.-L. H.,  M. N. and F. L. contributed equally to this work.
\end{acknowledgments}

\bibliographystyle{apsrev4-1}
\bibliography{references}

\end{document}


\title{Supplemental Material: \tim{Emulating quantum teleportation of a Majorana zero mode qubit} }

\author{He-Liang Huang}
\thanks{These three authors contributed equally}
\affiliation{Hefei National Laboratory for Physical Sciences at the Microscale and Department of Modern Physics, University of Science and Technology of China, Hefei 230026, China}
\affiliation{Shanghai Branch, CAS Center for Excellence in Quantum Information and Quantum Physics, University of Science and Technology of China, Shanghai 201315, China}
\affiliation{Shanghai Research Center for Quantum Sciences, Shanghai 201315, China}

\author{Marek Naro$\dot{\text{z}}$niak}
\thanks{These three authors contributed equally}
\affiliation{New York University Shanghai, 1555 Century Ave, Pudong, Shanghai 200122, China}
\affiliation{Department of Physics, New York University, New York, NY 10003, USA}

\author{Futian Liang}
\thanks{These three authors contributed equally}
\affiliation{Hefei National Laboratory for Physical Sciences at the Microscale and Department of Modern Physics, University of Science and Technology of China, Hefei 230026, China}
\affiliation{Shanghai Branch, CAS Center for Excellence in Quantum Information and Quantum Physics, University of Science and Technology of China, Shanghai 201315, China}
\affiliation{Shanghai Research Center for Quantum Sciences, Shanghai 201315, China}

\author{Youwei Zhao}
\author{Anthony D. Castellano}
\author{Ming Gong}
\author{Yulin Wu}
\author{Shiyu Wang}
\author{Jin Lin}
\author{Yu Xu}
\author{Hui Deng}
\author{Hao Rong}
\affiliation{Hefei National Laboratory for Physical Sciences at the Microscale and Department of Modern Physics, University of Science and Technology of China, Hefei 230026, China}
\affiliation{Shanghai Branch, CAS Center for Excellence in Quantum Information and Quantum Physics, University of Science and Technology of China, Shanghai 201315, China}
\affiliation{Shanghai Research Center for Quantum Sciences, Shanghai 201315, China}

\author{Jonathan P. Dowling}
\affiliation{Hearne Institute for Theoretical Physics, Department of Physics and Astronomy, Louisiana State University, Baton Rouge, Louisiana 70803, USA}
\affiliation{Hefei National Laboratory for Physical Sciences at Microscale and Department of Modern Physics,\\
University of Science and Technology of China, Hefei, Anhui 230026, China}
\affiliation{NYU-ECNU Institute of Physics at NYU Shanghai, 3663 Zhongshan Road North, Shanghai 200062, China}

\author{Cheng-Zhi Peng}

\affiliation{Hefei National Laboratory for Physical Sciences at the Microscale and Department of Modern Physics, University of Science and Technology of China, Hefei 230026, China}
\affiliation{Shanghai Branch, CAS Center for Excellence in Quantum Information and Quantum Physics, University of Science and Technology of China, Shanghai 201315, China}
\affiliation{Shanghai Research Center for Quantum Sciences, Shanghai 201315, China}

\author{Tim Byrnes}
\affiliation{New York University Shanghai, 1555 Century Ave, Pudong, Shanghai 200122, China}
\affiliation{NYU-ECNU Institute of Physics at NYU Shanghai, 3663 Zhongshan Road North, Shanghai 200062, China}
\affiliation{State Key Laboratory of Precision Spectroscopy, School of Physical and Material Sciences,East China Normal University, Shanghai 200062, China}
\affiliation{Department of Physics, New York University, New York, NY 10003, USA}

\author{Xiaobo Zhu}
\email{xbzhu16@ustc.edu.cn}
\author{Jian-Wei Pan}
\affiliation{Hefei National Laboratory for Physical Sciences at the Microscale and Department of Modern Physics, University of Science and Technology of China, Hefei 230026, China}
\affiliation{Shanghai Branch, CAS Center for Excellence in Quantum Information and Quantum Physics, University of Science and Technology of China, Shanghai 201315, China}
\affiliation{Shanghai Research Center for Quantum Sciences, Shanghai 201315, China}

\date{\today}

\setcounter{figure}{0}
\setcounter{equation}{0}
\setcounter{table}{0}
\makeatletter
\renewcommand{\thefigure}{S\@arabic\c@figure}
\renewcommand{\theequation}{S\@arabic\c@equation}
\renewcommand{\thetable}{S\@arabic\c@table}

\newcommand{\manuallabel}[2]{\def\@currentlabel{#2}\label{#1}}
\makeatother
\manuallabel{fig1}{1}
\manuallabel{fig2}{2}
\manuallabel{fig3}{3}

\maketitle

\section{Majorana Modes in the Kitaev Chain}

In this section we provide a brief review of Majorana modes in the Kitaev chain.  We refer the reader to several excellent reviews for further details \cite{leijnse2012introduction,preskill1998fault,nayak2008non,pachos2012introduction,lahtinen2017short}.

\subsection{Definition of Majorana fermions}

Consider a set of $ N $ fermions, which can be described standard fermionic anticommutation relations
\begin{align}
 \{ c_n, c_{n'}  \} & = 0 \nonumber \\
    \{ c_n, c_{n'}^\dagger \} & = \delta_{n n'},
\end{align}
where $ \delta_{n n'} $ is the Kronecker delta.
We may rewrite the operators for creating and annihilating a fermion on site $ n $ in terms of two \textit{Majorana fermions} in following way
\begin{align}
c_n &= \frac{1}{2}(\gamma_{n,\ell} + i\gamma_{n,r}) \\
c^\dagger_n &= \frac{1}{2}(\gamma_{n,\ell} - i\gamma_{n,r}) .
\label{fermionoperator}
\end{align}
These equations can be solved for $\gamma_{\ell}$ and $\gamma_r$ resulting with definitions of Majorana fermions in terms of a single fermion
\begin{align}
\gamma_{n,\ell} &= c_n + c^\dagger_n \\
\gamma_{n,r} &= -i c_n + i c^\dagger_n .
\end{align}
According to the definition, Majorana fermions are purely real
\begin{align}
\gamma_{n,\sigma} &= \gamma^\dagger_{n,\sigma},
\end{align}
where $ \sigma \in \{ \ell, r \}$.  They share similarities with standard fermions with regard to their anti-commutation property
\begin{align}
\{ \gamma_{n,\sigma}, \gamma_{n',\sigma'} \} = 2 \delta_{nn'} \delta_{\sigma \sigma'} \label{anti_commutation_majoranas} .
\end{align}
However, unlike standard fermions which obey the Pauli exclusion principle $ c_n^2 = (c_n^\dagger)^2 = 0 $, Majorana fermions are their own anti-particle and we have
\begin{align}
\gamma^2_{n,\sigma} &= 1 .
\label{squared_majoranas}
\end{align}

\subsection{Delocalized fermions}

Under this formalism, it appears the concept of Majorana fermions is just an algebraic manipulation. The interesting aspect of utilizing the Majorana operators arises when we construct other types of fermions that are not necessarily the physical fermions $ c_n $. Following the form of the fermion operators shown in (\ref{fermionoperator}), new {\it delocalized fermions}  can be defined using any pair of Majorana modes
\begin{align}
f_p &= \frac{1}{2}(\gamma_{n, \sigma} + i\gamma_{m,\nu}) \\
f_p^\dagger &= \frac{1}{2}(\gamma_{n, \sigma} - i\gamma_{m,\nu}) ,
\label{pseudofermions}
\end{align}
where $ \sigma, \nu \in \{ \ell, r \}$. Here
\begin{align}
    p \rightarrow (n,\sigma,m,\nu)
    \label{pairinglabel}
\end{align}
is a pairing label between two Majorana modes labeled by $(n,\sigma)$ and $ (m,\nu)$.
The fermion operator $ f_p $ is constructed from two Majorana modes, which are potentially at different physical sites $ n \ne m $, hence we call this a {\it delocalized fermion}.  A particular pair $ p $ always involves two different Majorana modes, such that $(n,\sigma) \ne  (m,\nu)$, meaning that a pair with both $ n= m $ and $ \sigma = \nu $ is not allowed.  It is possible however to have a pairing such that $ n = m $ but $ \sigma \ne \nu $, or $ n \ne m$ but $ \sigma = \nu$.  The former is exactly the case of physical fermions as shown in (\ref{fermionoperator}).

\begin{figure}
\begin{center}
\includegraphics[width=0.8\linewidth]{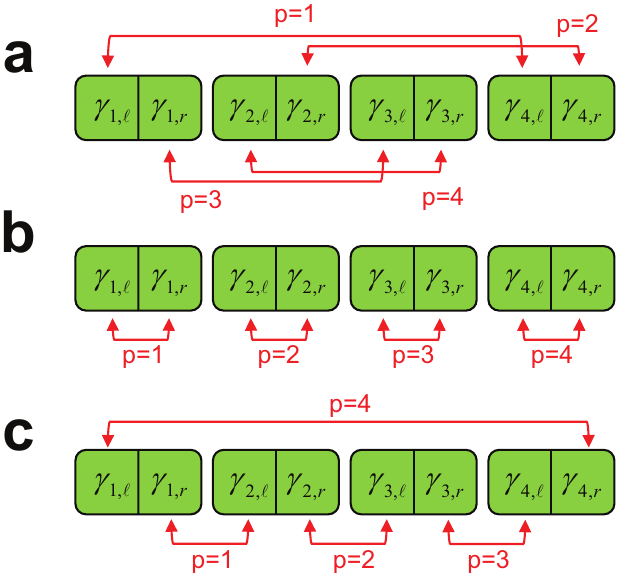}
\end{center}
\caption{Various Majorana mode pairing configurations.  (a) Random, (b) topologically trivial, and (c) Kitaev chain pairings are shown.
\label{sfig1}}
\end{figure}

Given a set of $ N $ fermions, and hence $ 2N $ Majorana modes, let us fix a particular pairing configuration labeled by (\ref{pairinglabel}).  Various examples of Majorana pairings are shown in Fig. \ref{sfig1}.  When establishing a pairing configuration, Majorana modes are never used twice, such that for different pairs $ p \ne p'$, the underlying Majoranas are all different.  Under these conditions, the anticommutation relations of the delocalized fermions (\ref{pseudofermions}) can be evaluated as
\begin{align}
\{f_p, f_{p'} \} &= \frac{1}{2} \left(
\delta_{n n'} \delta_{\sigma \sigma'}
+ i \delta_{n m'} \delta_{\sigma \nu'}
+ i \delta_{n' m} \delta_{\sigma' \nu}
-\delta_{m m'} \delta_{\nu \nu'} \right) \nonumber \\
& = 0
\label{fanticomm}
\end{align}
where the pairing label $ p' \rightarrow (n',\sigma',m',\nu') $.  The Kronecker delta functions simplify in (\ref{fanticomm}) because if $ p= p'$ it implies that
\begin{align}
    (n,\sigma) = (n',\sigma') \ne (m, \nu) = (m', \nu') ,
\end{align}
but if $ p \ne p '$ then it implies that
\begin{align}
    (n,\sigma) \ne (n',\sigma') \ne (m, \nu) \ne (m', \nu') .
\end{align}
Similarly we can evaluate
\begin{align}
\{f_p, f_{p'}^\dagger \} &= \frac{1}{2} \left(
\delta_{n n'} \delta_{\sigma \sigma'}
- i \delta_{n m'} \delta_{\sigma \nu'}
+ i \delta_{n' m} \delta_{\sigma' \nu}
+\delta_{m m'} \delta_{\nu \nu'} \right) \nonumber \\
& = \delta_{p p'}
\end{align}
which shows that the delocalized fermions are fermion operators as claimed.

\begin{figure}
\begin{center}
\includegraphics[width=0.8\linewidth]{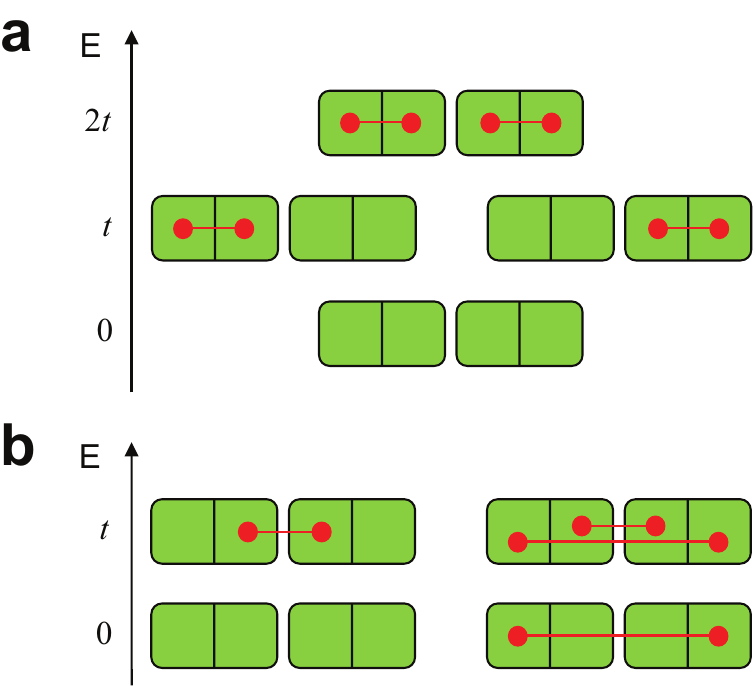}
\end{center}
\caption{Energy spectrum of Majorana pairing Hamiltonian for $ N = 2 $ fermions.   (a) Topologically trivial Hamiltonian (\ref{trivialham}) as shown in Fig. \ref{sfig1}(b).  (b) Kitaev Hamiltonian (\ref{kitaevham}) as shown in Fig. \ref{sfig1}(c).  The Majorana mode labels are suppressed, and the occupancy of the Majorana modes are denoted by the red bars.
\label{sfig2}}
\end{figure}

\subsection{Majorana pairing Hamiltonian}

To enforce a particular pairing configuration of Majorana modes, we must energetically stabilize the fermions that are defined by (\ref{pseudofermions}). For example, the Hamiltonian to enforce the regular fermion pairing (Fig. \ref{sfig1}(b)) is given by
\begin{align}
H &= t \sum_{n=1}^N  c^\dagger_n  c_n  \\
  &= \frac{t}{2} \sum_{n=1}^N (1 + i\gamma_{n, \ell} \gamma_{n, r}) ,
  \label{trivialham}
\end{align}
where $ t $ is some energy constant. The eigenstates of this Hamiltonian are given by
\begin{align}
    | j_1, \dots, j_N \rangle = \prod_{n=1}^N (c^\dagger_n)^{j_n} |0 \rangle
\end{align}
where $ j_n \in \{0,1 \} $ labels the occupancy of the $ n$th fermion. The energies of these states are
\begin{align}
    E = t  \sum_{n=1}^N j_n .
\end{align}
This can be rewritten in the Majorana language, where a fermion occupancy of the $ n $th site means that the underlying Majorana modes are both occupied.  Fig. \ref{sfig2}(a) shows the spectrum for the example of $ N =2$.

Similarly, for the delocalized fermions we can define a Majorana pairing Hamiltonian according to
\begin{align}
    H & = t \sum_{p=1}^N f^\dagger_p f_p,
    \label{fham}
\end{align}
where $ p $ runs over all $N $ Majorana pairs, for example that defined in Fig. \ref{sfig1}(a).  The eigenstates are again defined by the occupancy of the new fermions
\begin{align}
    | j_1, \dots, j_N \rangle = \prod_{p=1}^N (f^\dagger_p)^{j_p} |0 \rangle,
\end{align}
where $ | 0 \rangle $ is the ground state of (\ref{fham}), and $ j_p \in \{0,1 \}$ labels the occupancy of the $ p$th pair. The energy spectrum is again given by
\begin{align}
    E =  t  \sum_{p=1}^N j_p .
\end{align}

The Kitaev chain \cite{kitaev2003fault} is a particular example for the pairing configuration given in Fig. \ref{sfig1}(c).  In this case, the fermions are defined as
\begin{align}
    f_n = \frac{1}{2} ( \gamma_{n,r} + i \gamma_{n+1, \ell} ),
    \label{kitaevferm}
\end{align}
where $ n \in [1, N-1 ] $ for this operator.  Here a Majorana mode in the right box on site $ n $ is paired with another in the left box of site $ n + 1 $.  The Kitaev chain Hamiltonian is then
\begin{align}
    H & = t \sum_{n=1}^{N-1} f_n^\dagger f_n \nonumber \\
    & =  \frac{t}{2} \sum_{n=1}^{N-1} (1 + i\gamma_{n,r} \gamma_{n+1, \ell}) .
    \label{kitaevham}
\end{align}
Importantly, this Hamiltonian does {\it not} involve the Majorana pairing of the delocalized fermion corresponding to
\begin{align}
    f_N = \frac{1}{2} ( \gamma_{1,\ell} + i \gamma_{N, r} ) .
    \label{edgefermion}
\end{align}
This means that this fermion costs zero energy to excite, and makes every state in the spectrum of (\ref{kitaevham}) doubly degenerate, including the ground state.  As before, the eigenstates of (\ref{kitaevham}) are
\begin{align}
        | j_1, \dots, j_N \rangle = \prod_{n=1}^N (f^\dagger_n)^{j_n} |0 \rangle,
\end{align}
and the energy spectrum is
\textbf{\begin{align}
    E =  t  \sum_{n=1}^{N-1} j_n .
\end{align}}
This does not involve the occupancy label $ j_N $, which explicitly shows the double degeneracy.  An example of the spectrum of the Kitaev Hamiltonian is shown in Fig. \ref{sfig2}(b).

\subsection{Majorana zero modes}

The doubly degenerate ground states of the Kitaev Hamiltonian (\ref{kitaevham}) have zero energy and form a pair of orthogonal states. Let us as usual take the ground state with the absence of any fermions (\ref{kitaevferm}) or (\ref{edgefermion}) by  $\ket{0}$.  Then fermion operator (\ref{edgefermion}) then transforms this ground state into its degenerate pair, where the Majoranas on the end of the chain are occupied $ f^\dagger_N | 0 \rangle $.  These two states are used as the logical qubit states of the quantum computation, where
\begin{align}
    |0_L \rangle & \equiv | 0 \rangle   \nonumber \\
    |1_L \rangle & \equiv f^\dagger_N | 0 \rangle.
		\label{logicalstatesmzm}
\end{align}
For simplicity we denote in the main text $ f = f_N $, which is the annihilation operator for the edge states of the Kitaev Hamiltonian.  The Majorana modes $ \gamma_{1, \ell} $ and $ \gamma_{N,r} $  have zero energy and hence are called Majorana zero modes (MZMs).

One Kitaev chain therefore encodes one logical qubit's worth of information.  In order to have multiple logical qubits, then multiple Kitaev chains are required.  Labelling the Majorana mode labelled by $ (n,\sigma)$ on the $ m$th chain as $ \gamma_{n,\sigma}^{(m)}$, the  Hamiltonian for the multiple chain case then reads
\begin{align}
        H   = & \frac{t}{2} \sum_{n=1}^{N-1} (1 + i\gamma_{n,r}^{(m)} \gamma_{n+1, \ell}^{(m)})  \nonumber \\
 = & \frac{t}{2} \sum_{n=1}^{N-1}  ( 1 - {c^{(m)}_{n+1}}^\dagger c_n^{(m)} - {c_n^{(m)}}^\dagger c_{n+1}^{(m)} \nonumber \\
& + c_n^{(m)} c_{n+1}^{(m)}  + {c_{n+1}^{(m)}}^\dagger {c_n^{(m)}}^\dagger   ),
\end{align}
which up to a constant energy offset is the Hamiltonian (1) in the main text.  Here we denote the fermion annihilation  operator on the $n$th site of the $ m$th Kitaev chain as $ c_n^{(m)} $.

The MZMs on the $ m$th Kitaev chain then occur on the first and last Majorana sites and we define the operator
\begin{align}
    f^{(m)} = \frac{1}{2} ( \gamma_{1,\ell}^{(m)} + i  \gamma_{N,r}^{(m)} )
\end{align}
which destroys a MZM on the $ m$th chain. Here henceforth use the notation
\begin{align}
     \gamma_{\ell}^{(m)} & \equiv \gamma_{1,\ell}^{(m)} \nonumber \\
     \gamma_{r}^{(m)} & \equiv \gamma_{N,r}^{(m)} .
\end{align}
The full set of $ 2^M $ logical states are built up by applying the creation operator $ {f^{(m)}}^\dagger $ on the ground state $ |0\rangle $ contain zero Majorana modes.

\section{Braiding Majorana zero modes}

The MZMs are an example of non-Abelian anyons because their interchange causes a non-trivial effect on the ground state manifold.  In this section we derive the effect of braiding of the Majorana zero modes on two Kitaev chains.  It is sufficient to consider two Kitaev chains because we will consider the braiding of two MZMs to be the elementary process.  The two MZMs can originate from the same Kitaev chain, or one MZM each from two Kitaev chains.  This gives a total of 6 possible braidings of two MZMs, since each chain has two MZMs.

\subsection{Braiding operator}

Braiding two zero modes $\gamma_{n,\sigma}$ and $\gamma_{m,\nu}$ in a clockwise direction can be achieved by applying the operator \cite{leijnse2012introduction,nayak19962n,ivanov2001non}
\begin{align}
B_{(n,\sigma)  (m, \nu) } &= e^{\frac{\pi}{4} \gamma_{n,\sigma}  \gamma_{m,\nu} }= \frac{1}{\sqrt{2}} ( 1+ \gamma_{n,\sigma}  \gamma_{m,\nu} ) .     \label{operator_braiding}
\end{align}
It is apparent that this performs a braiding operation via the transformation
\begin{align}
  &   B_{(n,\sigma) (m, \nu) } \gamma_{n,\sigma} B_{(n,\sigma) (m, \nu) }^\dagger =  -  \gamma_{m,\nu} \nonumber \\
  & B_{(n,\sigma)  (m, \nu) } \gamma_{m,\nu} B_{(n,\sigma)   (m, \nu) }^\dagger = \gamma_{n,\sigma} .
\end{align}

When applied on the ground state manifold of the Kitaev chains, the braiding operators realize unitary operations on the MZM states.  Consider for the purposes of this section that there are $ M = 2 $ Kitaev chains, such that the logical states are
\begin{align}
    |00_L \rangle & \equiv |0 \rangle \nonumber \\
   |10_L \rangle & \equiv {f^{(1)}}^\dagger |0 \rangle \nonumber \\
     |01_L \rangle & \equiv {f^{(2)}}^\dagger  |0 \rangle \nonumber \\
      |11_L \rangle & \equiv  {f^{(1)}}^\dagger {f^{(2)}}^\dagger  |0 \rangle  ,
      \label{logical2qubit}
\end{align}
where $ | 0 \rangle $ is again the state with zero Majorana modes everywhere.  The purpose of the following section will be to derive the effect of various braiding operators on the logical space of states (\ref{logical2qubit}).

\begin{figure*}
\includegraphics[width=\textwidth]{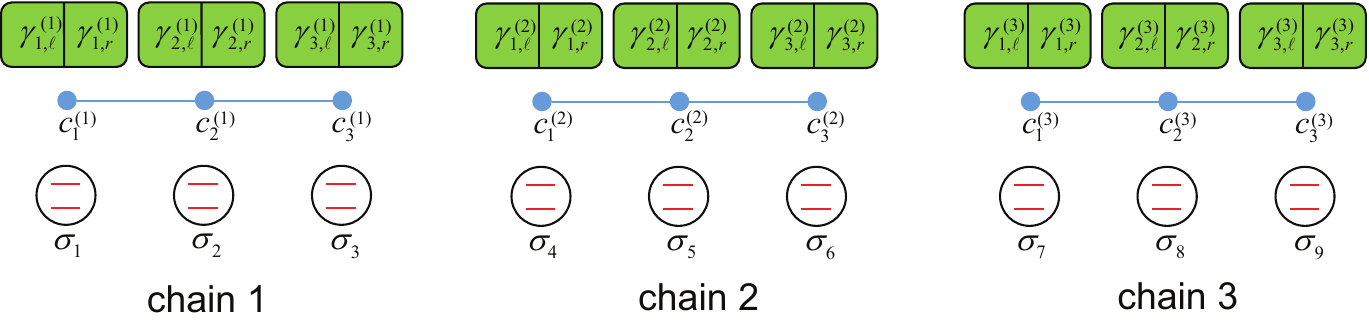}
\caption{Example of the labelling convention between Majorana modes, fermions, and spin operators for the case of $ M = 3 $ chains each with $ N = 3$ fermions.
\label{sfig3}}
\end{figure*}

\subsection{Spin representation}

For each braiding operator acting on logical space, an equivalent spin operator can be derived acting on corresponding physical space.  This is done using Jordan-Wigner transformation to transform the Majorana variables to spin variables.  We consider a layout of spins as shown in Fig. \ref{sfig3}. The $M$ Kitaev chains each with $N$ fermions are arranged in a larger chains, in ascending order.  We label the spin operators from 1 to $ NM$, the total number of fermions and spins in the mapping.  In this case, the MZM can be transformed to spin variables according to
\begin{align}
\gamma_{\ell}^{(p)} &=  \left( \prod^{pN-N}_{i=1} \sigma^z_i\right) \sigma^x_{pN-N+1} \nonumber  \\
\gamma_{r}^{(p)} &=  \left( \prod^{pN-1}_{i=1} \sigma^z_i \right)  \sigma^y_{pN},
\end{align}
where $p$ is the chain index. In the calculations below, we only consider two chains, and hence it is convenient to explicitly write the spin mapped MZM operators
\begin{align}
    \gamma^{(1)}_{\ell} & = \sigma^x_1 \nonumber \\
\gamma^{(1)}_r & = \left( \prod_{k=1}^{N-1} \sigma^z_k  \right) \sigma^y_N
= \sigma^z_1 \dots \sigma^z_{N-1}  \sigma^y_N  \nonumber \\
\gamma^{(2)}_{\ell} & = \left( \prod_{k=1}^{N} \sigma^z_k  \right)   \sigma^x_{N+1}  = \sigma^z_1 \dots \sigma^z_{N}  \sigma^x_{N+1}   \nonumber \\
\gamma^{(2)}_r &  = \left( \prod_{k=1}^{2N-1} \sigma^z_k  \right) \sigma^y_{2N} = \sigma^z_1 \dots \sigma^z_{2N-1}  \sigma^y_{2N}  \label{jwtrans}
\end{align}

The logical states on which Jordan-Wigner transformed operators act are defined explicitly in a following way
\begin{align}
    \ket{0_L} &= \frac{1}{\sqrt{2}}(\ket{+\dots+}+\ket{-\dots-})\\
    \ket{1_L} &= \frac{1}{\sqrt{2}}(\ket{+\dots+}-\ket{-\dots-})
    \label{logicalstates}
\end{align}

\subsection{Derivation of the six braiding gates in Fig.~\ref{fig2new}}

\subsubsection{$ \gamma_{\ell}^{(1)} \rightleftarrows  \gamma_{r}^{(1)}  $ braid:  $\sqrt{Z_1}$ gate}

We can express this braiding operation in terms of spin operators by applying Jordan-Wigner transformation (\ref{jwtrans})
\begin{align}
B_{(1,\ell), (1,r)} &= e^{\frac{\pi}{4}\gamma_{\ell}^{(1)} \gamma_{r}^{(1)}} \\
&= \frac{1}{\sqrt{2}}(1 + \gamma_{\ell}^{(1)} \gamma_{r}^{(1)}) \\
&= \frac{1}{\sqrt{2}}(1 + \sigma^x_1 \sigma^z_1 \dots  \sigma^z_{N-1}    \sigma^y_N)  \\
&= \frac{1}{\sqrt{2}}(1 - i \sigma^y_1 \sigma^z_2 \dots \sigma^z_{N-1}   \sigma^y_N) . \label{physicalsqrtz1}
\end{align}
This is the braiding operator in the spin representation.

To see the effect of this braiding operator in the logical space, we operate the above state on the spin representation of the logical states (\ref{logicalstates}).  For compactness let us first define the non-trivial part of the braiding operator as
\begin{align}
    \Gamma = - i \sigma^y_1 \sigma^z_2 \dots \sigma^z_{N-1}   \sigma^y_N .
\end{align}
Applying $ \Gamma $ on the logical states we find
\begin{align}
\Gamma \ket{0_L} = & \frac{i}{\sqrt{2}}(\ket{-\dots-} + \ket{+\dots+}) \\
 = & i \ket{0_L}
\end{align}
and
\begin{align}
\Gamma  \ket{1_L} =  & -\frac{i}{\sqrt{2}}(\ket{+\dots+} - \ket{-\dots-}) \\
 = & -i\ket{1_L} .
\end{align}
Here we used the fact that
\begin{align}
    \sigma^x | \pm \rangle & = \pm | \pm \rangle \nonumber \\
    \sigma^y | \pm \rangle & = \mp i| \mp \rangle \nonumber \\
        \sigma^z | \pm \rangle & = | \mp \rangle .
        \label{paulirelations}
\end{align}
Since we can write
\begin{align}
    B_{(1,\ell), (1,r)}  =\frac{1}{\sqrt{2}}(1 + \Gamma) ,
\end{align}
it then follows that
\begin{align}
    B_{(1,\ell), (1,r)} | 0_L \rangle &  = \frac{1}{\sqrt{2}}(1 + i ) | 0_L \rangle = e^{i \pi/4} | 0_L \rangle \nonumber \\
     B_{(1,\ell), (1,r)} |1_L \rangle &  = \frac{1}{\sqrt{2}}(1 - i ) | 1_L \rangle  = e^{-i \pi/4} | 0_L \rangle .
\end{align}
This corresponds to the $\sqrt{Z_1}$ operator.

\subsubsection{$ \gamma_{\ell}^{(1)} \rightleftarrows  \gamma_{\ell}^{(2)}  $ braid: $\sqrt{Y_1 X_2}$ gate}

For braiding involving more than one chain we apply the same method, substituting the Jordan-Wigner transformation (\ref{jwtrans}) into the braiding operator
\begin{align}
B_{(1,\ell), (2,\ell)} &= e^{\frac{\pi}{4}\gamma_{\ell}^{(1)} \gamma_{\ell}^{(2)}}  \nonumber \\
&= \frac{1}{\sqrt{2}}(1 + \gamma_{\ell}^{(1)} \gamma_{\ell}^{(2)}) \nonumber  \\
&= \frac{1}{\sqrt{2}}(1 + \sigma^x_1 \sigma^z_1 \dots \sigma^z_N \sigma^x_{N+1})  \nonumber \\
&= \frac{1}{\sqrt{2}}(1 - i \sigma^y_1 \sigma^z_1 \dots \sigma^z_N \sigma^x_{N+1}) \label{physicalsqrtyx}
\end{align}
This is the braiding operator in the spin representation.

To examine the effect on the logical states, we again define the non-trivial part of the above operator as
\begin{align}
    \Gamma = - i \sigma^y_1 \sigma^z_1 \dots \sigma^z_N \sigma^x_{N+1} .
\end{align}
Using the relations (\ref{paulirelations}), we can evaluate
\begin{align}
    \Gamma | 00_L \rangle & = | 11_L \rangle  \nonumber \\
     \Gamma | 01_L \rangle & = | 10_L \rangle  \nonumber \\
         \Gamma | 10_L \rangle & = - | 01_L \rangle  \nonumber \\
         \Gamma | 11_L \rangle & = - | 00_L \rangle  .
\end{align}
where we used the explicit expansions
\begin{align}
\ket{00_L} =
    &\begin{aligned}[t]
    \frac{1}{\sqrt{2}}(&\ket{+\dots+}\ket{+\dots+} + \ket{+\dots+}\ket{-\dots-} \nonumber \\
    +&\ket{-\dots-}\ket{+\dots+}  \ket{-\dots-}\ket{-\dots-})  \nonumber \\
    \end{aligned}  \nonumber \\
\ket{01_L} =
    &\begin{aligned}[t]
    \frac{1}{\sqrt{2}}(&\ket{+\dots+}\ket{+\dots+}-\ket{+\dots+}\ket{-\dots-} \\
    +&\ket{-\dots-}\ket{+\dots+} - \ket{-\dots-}\ket{-\dots-}) \\
    \end{aligned} \nonumber \\
\ket{10_L} =
    &\begin{aligned}[t]
    \frac{1}{\sqrt{2}}(&\ket{+\dots+}\ket{+\dots+}+ \ket{+\dots+}\ket{-\dots-} \\
    -& \ket{-\dots-}\ket{+\dots+} - \ket{-\dots-}\ket{-\dots-})  \\
    \end{aligned} \nonumber \\
\ket{11_L} =
    &\begin{aligned}[t]
    \frac{1}{\sqrt{2}}(&\ket{+\dots+}\ket{+\dots+} - \ket{+\dots+}\ket{-\dots-} \\
    -&\ket{-\dots-}\ket{+\dots+} + \ket{-\dots-}\ket{-\dots-}) \\
    \end{aligned}  .\label{logical_11}
\end{align}
Applying the braiding operator
\begin{align}
    B_{(1,\ell), (2,\ell)} = \frac{1}{\sqrt{2}}( 1+ \Gamma)
\end{align}
then gives
\begin{align}
B_{(1,\ell), (2,\ell)} | 00_L \rangle & = \frac{1}{\sqrt{2}}
( \ket{00_L} + \ket{11_L} )  \nonumber \\
B_{(1,\ell), (2,\ell)} | 01_L \rangle & = \frac{1}{\sqrt{2}}
( \ket{01_L} + \ket{10_L} )  \nonumber \\
B_{(1,\ell), (2,\ell)} | 10_L \rangle & = \frac{1}{\sqrt{2}}
( \ket{10_L} - \ket{01_L} )  \nonumber \\
B_{(1,\ell), (2,\ell)} | 11_L \rangle & = \frac{1}{\sqrt{2}}
( \ket{11_L} - \ket{00_L} )  .
\end{align}
This corresponds to the $ \sqrt{Y_1 X_2} $ gate.

\subsubsection{$ \gamma_{\ell}^{(1)} \rightleftarrows  \gamma_{r}^{(2)}  $ braid:  $\sqrt{Y_1 Y_2}$ gate}

\begin{align}
B^\prime_{(1,\ell), (2,r)} &= e^{\frac{\pi}{4}\gamma_{\ell}^{(1)} \gamma_{r}^{(2)}} \\
&= \frac{1}{\sqrt{2}}(1 + \gamma_{\ell}^{(1)} \gamma_{r}^{(2)}) \\
&= \frac{1}{\sqrt{2}}(1 + \sigma^x_1 \sigma^z_{1} \dots \sigma^z_{N} \sigma^z_{1} \dots \sigma^z_{2N-1} \sigma^x_{2N}) \\
&= \frac{1}{\sqrt{2}}(1 -i \sigma^y_1 \sigma^z_{N+1} \dots \sigma^z_{2N-1} \sigma^x_{2N}) \label{braid_yy_spin_rep} \\
&= \frac{1}{\sqrt{2}}(1 + B^\prime_{(1,\ell), (2,r)})
\end{align}

We demonstrate the correctness of this operator applied to the logical space and replicating the following.

\begin{align}
\sqrt{Y_1 Y_2} \ket{00_L} &= \frac{1}{\sqrt{2}} ( \ket{00_L} + i\ket{11_L} ) \label{logic_yy_00} \\
\sqrt{Y_1 Y_2} \ket{01_L} &= \frac{1}{\sqrt{2}} ( \ket{01_L} - i\ket{10_L} ) \\
\sqrt{Y_1 Y_2} \ket{10_L} &= \frac{1}{\sqrt{2}} ( \ket{10_L} - i\ket{01_L} ) \\
\sqrt{Y_1 Y_2} \ket{11_L} &= \frac{1}{\sqrt{2}} ( \ket{11_L} - i\ket{00_L} ) \label{logic_yy_11}
\end{align}

Now we apply $B^\prime_{(1,\ell), (2,r)}$ to states (\ref{logical_11}) and we replicate (\ref{logic_yy_00}-\ref{logic_yy_11}) as follows.

Let $B=B_{(1,\ell), (2,r)}$ and $B^\prime=-i \sigma^y_1 \sigma^z_{N+1} \dots \sigma^z_{2N-1} \sigma^x_{2N})$

\begin{align}
B \ket{00_L} = \frac{1}{\sqrt{2}}(
    &\begin{aligned}[t]
    \ket{00_L}
    +B^\prime(&\ket{+\dots+}\ket{+\dots+} \\
    +&\ket{+\dots+}\ket{-\dots-}) \\
    +&\ket{-\dots-}\ket{+\dots+}) \\
    +&\ket{-\dots-}\ket{-\dots-}) \\
    \end{aligned} \label{apply_yy_00} \\
= \frac{1}{\sqrt{2}}(
    &\begin{aligned}[t]
    \ket{00_L}
    -i(&\ket{-\dots-}\ket{-\dots-} \\
    -&\ket{-\dots-}\ket{+\dots+}) \\
    -&\ket{+\dots+}\ket{-\dots-}) \\
    +&\ket{+\dots+}\ket{+\dots+}) \\
    \end{aligned} \\
= & \frac{1}{\sqrt{2}}(\ket{00_L} + i\ket{11_L}) \\
B \ket{01_L} = \frac{1}{\sqrt{2}}(
    &\begin{aligned}[t]
    \ket{01_L}
    +B^\prime(&\ket{+\dots+}\ket{+\dots+} \\
    -&\ket{+\dots+}\ket{-\dots-}) \\
    +&\ket{-\dots-}\ket{+\dots+}) \\
    -&\ket{-\dots-}\ket{-\dots-}) \\
    \end{aligned} \label{apply_yy_01} \\
= \frac{1}{\sqrt{2}}(
    &\begin{aligned}[t]
    \ket{01_L}
    -i(&-\ket{-\dots-}\ket{-\dots-} \\
    -&\ket{-\dots-}\ket{+\dots+}) \\
    +&\ket{+\dots+}\ket{-\dots-}) \\
    +&\ket{+\dots+}\ket{+\dots+}) \\
    \end{aligned} \\
= & \frac{1}{\sqrt{2}}(\ket{01_L} - i\ket{10_L}) \\
B \ket{10_L} = \frac{1}{\sqrt{2}}(
    &\begin{aligned}[t]
    \ket{10_L}
    +B^\prime(&\ket{+\dots+}\ket{+\dots+} \\
    +&\ket{+\dots+}\ket{-\dots-}) \\
    -&\ket{-\dots-}\ket{+\dots+}) \\
    -&\ket{-\dots-}\ket{-\dots-}) \\
    \end{aligned} \label{apply_yy_10} \\
= \frac{1}{\sqrt{2}}(
    &\begin{aligned}[t]
    \ket{10_L}
    -i(&-\ket{-\dots-}\ket{-\dots-} \\
    +&\ket{-\dots-}\ket{+\dots+}) \\
    -&\ket{+\dots+}\ket{-\dots-}) \\
    +&\ket{+\dots+}\ket{+\dots+}) \\
    \end{aligned} \\
= & \frac{1}{\sqrt{2}}(\ket{10_L} - i \ket{01_L}) \\
B \ket{11_L} = \frac{1}{\sqrt{2}}(
    &\begin{aligned}[t]
    \ket{11_L}
    +B^\prime(&\ket{+\dots+}\ket{+\dots+} \\
    -&\ket{+\dots+}\ket{-\dots-}) \\
    -&\ket{-\dots-}\ket{+\dots+}) \\
    +&\ket{-\dots-}\ket{-\dots-}) \\
    \end{aligned} \label{apply_yy_11} \\
= \frac{1}{\sqrt{2}}(
    &\begin{aligned}[t]
    \ket{11_L}
    -i(&-\ket{-\dots-}\ket{-\dots-} \\
    -&\ket{-\dots-}\ket{+\dots+}) \\
    -&\ket{+\dots+}\ket{-\dots-}) \\
    -&\ket{+\dots+}\ket{+\dots+}) \\
    \end{aligned} \\
= & \frac{1}{\sqrt{2}}(\ket{11_L} +i \ket{00_L}) \\
\end{align}

\subsubsection{$ \gamma_{r}^{(1)} \rightleftarrows  \gamma_{\ell}^{(2)}  $ braid:  $\sqrt{X_1 X_2}$ gate}

\begin{align}
B^\prime_{(1,r), (2,\ell)} &= e^{\frac{\pi}{4}\gamma_{r}^{(1)} \gamma_{\ell}^{(2)}} \\
&= \frac{1}{\sqrt{2}}(1 + \gamma_{r}^{(1)} \gamma_{\ell}^{(2)}) \\
&= \frac{1}{\sqrt{2}}(1 + \sigma^z_{1} \dots \sigma^z_{N-1} \sigma^y_N \sigma^z_{1} \dots \sigma^z_{N} \sigma^x_{N+1}) \\
&= \frac{1}{\sqrt{2}}(1 + i \sigma^x_N \sigma^x_{N+1}) \label{braid_xx_spin_rep} \\
&= \frac{1}{\sqrt{2}}(1 + B^\prime_{(1,r), (2,\ell)})
\end{align}

We demonstrate the correctness of this operator applied to the logical space and replicating the following.

\begin{align}
\sqrt{X_1 X_2} \ket{00_L} &= \frac{1}{\sqrt{2}} ( \ket{00_L} + i\ket{11_L} ) \label{logic_xx_00} \\
\sqrt{X_1 X_2} \ket{01_L} &= \frac{1}{\sqrt{2}} ( \ket{01_L} + i\ket{10_L} ) \\
\sqrt{X_1 X_2} \ket{10_L} &= \frac{1}{\sqrt{2}} ( \ket{10_L} + i\ket{01_L} ) \\
\sqrt{X_1 X_2} \ket{11_L} &= \frac{1}{\sqrt{2}} ( \ket{11_L} + i\ket{00_L} ) \label{logic_xx_11}
\end{align}

Now we apply $B^\prime_{(1,r), (2,\ell)}$ to states (\ref{logical_11}) and we replicate (\ref{logic_xx_00}-\ref{logic_xx_11}) as follows.

Let $B=B_{(1,r), (2,\ell)}$ and $B^\prime=-i \sigma^x_N \sigma^x_{N+1}$

\begin{align}
B \ket{00_L} = \frac{1}{\sqrt{2}}(
    &\begin{aligned}[t]
    \ket{00_L}
    +B^\prime(&\ket{+\dots+}\ket{+\dots+} \\
    +&\ket{+\dots+}\ket{-\dots-}) \\
    +&\ket{-\dots-}\ket{+\dots+}) \\
    +&\ket{-\dots-}\ket{-\dots-}) \\
    \end{aligned} \label{apply_xx_00} \\
= \frac{1}{\sqrt{2}}(
    &\begin{aligned}[t]
    \ket{00_L}
    +i(&\ket{+\dots+}\ket{+\dots+} \\
    -&\ket{+\dots+}\ket{-\dots-}) \\
    -&\ket{-\dots-}\ket{+\dots+}) \\
    +&\ket{-\dots-}\ket{-\dots-}) \\
    \end{aligned} \\
= & \frac{1}{\sqrt{2}}(\ket{00_L} + i\ket{11_L}) \\
B \ket{01_L} = \frac{1}{\sqrt{2}}(
    &\begin{aligned}[t]
    +B^\prime(&\ket{+\dots+}\ket{+\dots+} \\
    +&\ket{+\dots+}\ket{-\dots-}) \\
    -&\ket{-\dots-}\ket{+\dots+}) \\
    -&\ket{-\dots-}\ket{-\dots-}) \\
    \end{aligned} \label{apply_xx_01} \\
= & \frac{1}{\sqrt{2}}(\ket{01_L} + i \ket{10_L}) \\
= \frac{1}{\sqrt{2}}(
    &\begin{aligned}[t]
    \ket{01_L}
    +i(&\ket{+\dots+}\ket{+\dots+} \\
    +&\ket{+\dots+}\ket{-\dots-}) \\
    -&\ket{-\dots-}\ket{+\dots+}) \\
    -&\ket{-\dots-}\ket{-\dots-}) \\
    \end{aligned} \\
= & \frac{1}{\sqrt{2}}(\ket{01_L} + i\ket{10_L}) \\
B \ket{10_L} = \frac{1}{\sqrt{2}}(
    &\begin{aligned}[t]
    \ket{10_L}
    +B^\prime(&\ket{+\dots+}\ket{+\dots+} \\
    +&\ket{+\dots+}\ket{-\dots-}) \\
    -&\ket{-\dots-}\ket{+\dots+}) \\
    -&\ket{-\dots-}\ket{-\dots-}) \\
    \end{aligned} \label{apply_xx_10} \\
= \frac{1}{\sqrt{2}}(
    &\begin{aligned}[t]
    \ket{10_L}
    +i(&\ket{+\dots+}\ket{+\dots+} \\
    -&\ket{+\dots+}\ket{-\dots-}) \\
    +&\ket{-\dots-}\ket{+\dots+}) \\
    -&\ket{-\dots-}\ket{-\dots-}) \\
    \end{aligned} \\
= & \frac{1}{\sqrt{2}}(\ket{10_L} + i \ket{01_L}) \\
B \ket{11_L} = \frac{1}{\sqrt{2}}(
    &\begin{aligned}[t]
    \ket{11_L}
    +B^\prime(&\ket{+\dots+}\ket{+\dots+} \\
    -&\ket{+\dots+}\ket{-\dots-}) \\
    -&\ket{-\dots-}\ket{+\dots+}) \\
    +&\ket{-\dots-}\ket{-\dots-}) \\
    \end{aligned} \label{apply_xx_11} \\
= \frac{1}{\sqrt{2}}(
    &\begin{aligned}[t]
    \ket{11_L}
    +i(&\ket{+\dots+}\ket{+\dots+} \\
    +&\ket{+\dots+}\ket{-\dots-}) \\
    +&\ket{-\dots-}\ket{+\dots+}) \\
    +&\ket{-\dots-}\ket{-\dots-}) \\
    \end{aligned} \\
= & \frac{1}{\sqrt{2}}(\ket{11_L} +i \ket{00_L}) \\
\end{align}

\subsubsection{$ \gamma_{r}^{(1)} \rightleftarrows  \gamma_{r}^{(2)}  $ braid:  $\sqrt{X_1 Y_2}$ gate}

\begin{align}
B^\prime_{(1,r), (2,r)} &= e^{\frac{\pi}{4}\gamma_{r}^{(1)} \gamma_{r}^{(2)}} \\
&= \frac{1}{\sqrt{2}}(1 + \gamma_{r}^{(1)} \gamma_{r}^{(2)}) \\
&= \frac{1}{\sqrt{2}}(1 + \sigma^y_N \sigma^z_{N} \dots \sigma^z_{2N-1} \sigma^x_{2N}) \\
&= \frac{1}{\sqrt{2}}(1 + i\sigma^x_1 \sigma^z_{N+1} \dots \sigma^z_{2N-1} \sigma^x_{2N}) \\
&= \frac{1}{\sqrt{2}}(1 + B^\prime_{(1,r), (2,r)})
\end{align}

We demonstrate the correctness of this operator applied to the logical space and replicating the following.

\begin{align}
\sqrt{X_1 Y_2} \ket{00_L} &= \frac{1}{\sqrt{2}} ( \ket{00_L} + \ket{11_L} ) \label{logic_xy_00} \\
\sqrt{X_1 Y_2} \ket{01_L} &= \frac{1}{\sqrt{2}} ( \ket{01_L} - \ket{10_L} ) \\
\sqrt{X_1 Y_2} \ket{10_L} &= \frac{1}{\sqrt{2}} ( \ket{10_L} + \ket{01_L} ) \\
\sqrt{X_1 Y_2} \ket{11_L} &= \frac{1}{\sqrt{2}} ( \ket{11_L} - \ket{00_L} ) \label{logic_xy_11}
\end{align}

Now we apply $B^\prime_{(1,r), (2,r)}$ to states (\ref{logical_11}) and we replicate (\ref{logic_xy_00}-\ref{logic_xy_11}) as follows.

Let $B=B_{(1,r), (2,r)}$ and $B^\prime=i\sigma^x_1 \sigma^z_{N+1} \dots \sigma^z_{2N-1} \sigma^x_{2N}$

\begin{align}
B \ket{00_L} = \frac{1}{\sqrt{2}}(
    &\begin{aligned}[t]
    \ket{00_L}
    +B^\prime(&\ket{+\dots+}\ket{+\dots+} \\
    +&\ket{+\dots+}\ket{-\dots-}) \\
    +&\ket{-\dots-}\ket{+\dots+}) \\
    +&\ket{-\dots-}\ket{-\dots-}) \\
    \end{aligned} \label{apply_xy_00} \\
= \frac{1}{\sqrt{2}}(
    &\begin{aligned}[t]
    \ket{00_L}
    -(&\ket{+\dots+}\ket{-\dots-} \\
    -&\ket{+\dots+}\ket{+\dots+}) \\
    -&\ket{-\dots-}\ket{-\dots-}) \\
    +&\ket{-\dots-}\ket{+\dots+}) \\
    \end{aligned} \\
= & \frac{1}{\sqrt{2}}(\ket{00_L} + \ket{11_L}) \\
B \ket{01_L} = \frac{1}{\sqrt{2}}(
    &\begin{aligned}[t]
    \ket{01_L}
    +B^\prime(&\ket{+\dots+}\ket{+\dots+} \\
    -&\ket{+\dots+}\ket{-\dots-}) \\
    +&\ket{-\dots-}\ket{+\dots+}) \\
    -&\ket{-\dots-}\ket{-\dots-}) \\
    \end{aligned} \label{apply_xy_01} \\
= \frac{1}{\sqrt{2}}(
    &\begin{aligned}[t]
    \ket{01_L}
    -(&\ket{+\dots+}\ket{-\dots-} \\
    +&\ket{+\dots+}\ket{+\dots+}) \\
    -&\ket{-\dots-}\ket{-\dots-}) \\
    -&\ket{-\dots-}\ket{+\dots+}) \\
    \end{aligned} \\
= & \frac{1}{\sqrt{2}}(\ket{01_L} - \ket{10_L}) \\
B \ket{10_L} = \frac{1}{\sqrt{2}}(
    &\begin{aligned}[t]
    \ket{10_L}
    +B^\prime(&\ket{+\dots+}\ket{+\dots+} \\
    +&\ket{+\dots+}\ket{-\dots-}) \\
    -&\ket{-\dots-}\ket{+\dots+}) \\
    -&\ket{-\dots-}\ket{-\dots-}) \\
    \end{aligned} \label{apply_xy_10} \\
= \frac{1}{\sqrt{2}}(
    &\begin{aligned}[t]
    \ket{10_L}
    -(-&\ket{+\dots+}\ket{-\dots-} \\
    +&\ket{+\dots+}\ket{+\dots+}) \\
    -&\ket{-\dots-}\ket{-\dots-}) \\
    +&\ket{-\dots-}\ket{+\dots+}) \\
    \end{aligned} \\
= & \frac{1}{\sqrt{2}}(\ket{10_L} + \ket{01_L}) \\
B \ket{11_L} = \frac{1}{\sqrt{2}}(
    &\begin{aligned}[t]
    \ket{11_L}
    +B^\prime(&\ket{+\dots+}\ket{+\dots+} \\
    -&\ket{+\dots+}\ket{-\dots-}) \\
    -&\ket{-\dots-}\ket{+\dots+}) \\
    +&\ket{-\dots-}\ket{-\dots-}) \\
    \end{aligned} \label{apply_xy_11} \\
= \frac{1}{\sqrt{2}}(
    &\begin{aligned}[t]
    \ket{11_L}
    -(&\ket{+\dots+}\ket{-\dots-} \\
    +&\ket{+\dots+}\ket{+\dots+}) \\
    +&\ket{-\dots-}\ket{-\dots-}) \\
    +&\ket{-\dots-}\ket{+\dots+}) \\
    \end{aligned} \\
= & \frac{1}{\sqrt{2}}(\ket{11_L} - \ket{00_L}) \\
\end{align}

\subsubsection{$ \gamma_{\ell}^{(2)} \rightleftarrows  \gamma_{r}^{(2)}  $ braid: $\sqrt{Z_2}$ gate}

The $\sqrt{Z_2}$ gate represents same braid as $\sqrt{Z_1}$ the difference is all indices are shifted by $N$ which is the length of single logical qubit.

\begin{figure*}[!htbp]
\begin{center}
\includegraphics[width=0.9\linewidth]{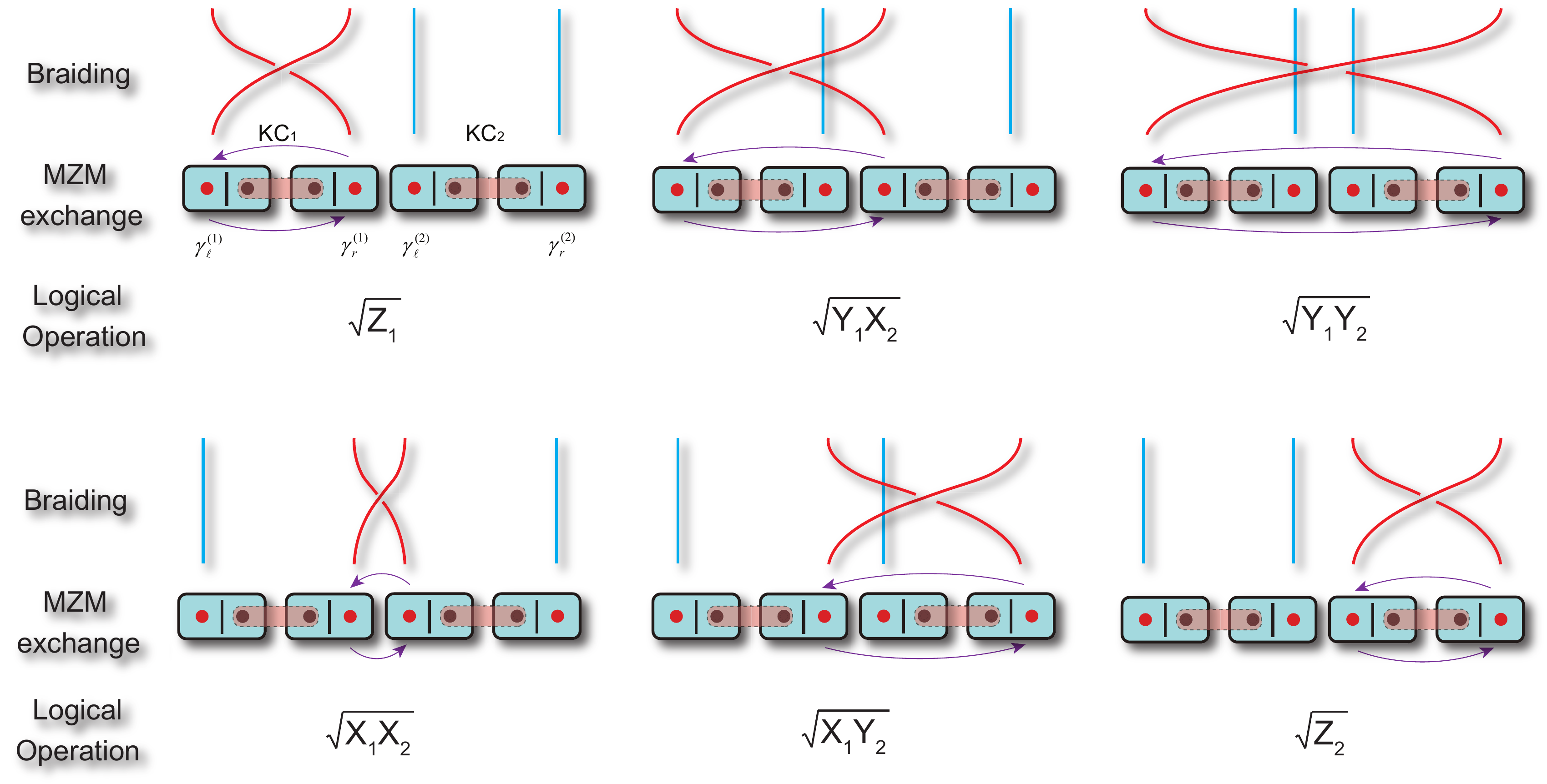}
\end{center}
\caption{Majorana modes and their braiding operations. The six possible braiding operations for two Kitaev chains (KC), and the effect in terms of the logical states. The left- and right-most Majorana Zero Mode (MZM)
on chains 1 and 2 are labeled by $ \gamma_{\ell,r}^{(1,2)} $ respectively.   We denote the Pauli operators for the underlying physical qubits by $ \sigma^{x},\sigma^{y},\sigma^{z} $ and the higher level logical operators by $ X, Y, Z $.
\label{fig2new}
}
\end{figure*}

\tim{\section{Error protection of Majorana Zero Modes in the Kitaev chain}

In this section we explain why the logical MZM states of the Kitaev chain are protected against errors, and the nature of the error protection after the Jordan-Wigner spin mapping.

To understand the nature of the error protection of the MZM states in the original fermion Hamiltonian (Eq. (1) in the main text), let us revisit Kitaev’s original paper \cite{kitaev2001unpaired} introducing the model.  In Ref. \cite{kitaev2001unpaired}, it is explained why information encoded in this system is protected.  The suggestion in this paper is to use delocalized fermion degrees of freedom (consisting of MZMs) as the logical qubits.  Specifically, these are given by (\ref{logicalstatesmzm}) in our notation.  It is then explained that in order to protect the quantum information, we must protect against both logical $ X $  and $ Z $ errors.  We look at both of these in turn.

For the logical $ X $  errors, in terms of the delocalized fermions, Majoranas, and original fermions, this can be written respectively as
\begin{align}
X = f_N + f_N^\dagger = \gamma_{1,\ell}  = c_1 + c_1^\dagger .
\end{align}
We see that in terms of the underlying fermions, such errors require either the loss or gain of a single fermion.  Such terms are charge non-conserving, and even under a superconducting Hamiltonian, are fermion parity non-conserving (since BCS terms have two fermion annihilation/creation operators).  This means that they are unlikely to naturally occur.  Meanwhile, in the spin version of the Hamiltonian, these types of errors {\it can} occur since after the Jordan-Wigner mapping we have
\begin{align}
X = c_1 + c_1^\dagger = \sigma_1^x .
\end{align}
In the transverse Ising model, this appears as a magnetic field in the $ x $-direction, which has no reason not to be present.  The lack of protection is also evident from Eq. (4) in the main text, where applying a $ \sigma_1^x $-error gives
\begin{align}
 \sigma_1^x | 0_L \rangle = \frac{1}{\sqrt{2}} ( | ++ \rangle - | -- \rangle ) = | 1_L \rangle ,
\end{align}
which is a logical error. We thus see that the spin mapping renders the logical bits susceptible to  $ X $ errors.

Now let us look at  $ Z $ errors.  Here we may evaluate
\begin{align}
Z & = 1 - 2 f^\dagger_N f_N \nonumber \\
& = c_1 c_N + c_N^\dagger c_1 + c_1^\dagger c_N + c^\dagger_N c^\dagger_1 .
\end{align}
The point made in Ref. \cite{kitaev2001unpaired} is that such a logical error consists of physically delocalized interactions between fermion sites 1 and $N$.  By well-separating the ends of the chain, we can suppress such logical errors, and hence suppress $ Z $ errors.  Looking at this in the spin language we have
\begin{align}
Z = \sigma^y_1 \sigma^z_2 \dots \sigma^z_{N-1} \sigma^y_N  .
\end{align}
This still possesses the delocalized form that is desired for error protection, since there must be an interaction between sites 1 and $N$.  In fact, it is arguably even better protected than the fermionic form, since it involves strings of $ \sigma^z_n $  operators arising from the Jordan-Wigner transformation.  Thus, we expect that for logical $ Z $ errors, the spin encoding should still offer protection.  In our case, physical $ \sigma^z_n $ errors can be detected and removed from the final results via postselection as explained in the main text. This results in an improved teleportation fidelity, as it removes single qubit phase errors that may occur during the teleportation circuit.
  }

\section{Modified teleportation circuit}
\label{sec:tele}  Our version of the teleportation circuit uses slightly different gates to the standard version of teleportation, such as that given in Ref. \cite{nielsen2002quantum}, where Hadamard and CNOT gates are used.  Instead of these gates, we based our teleportation on the $ \sqrt{X_1 X_2} $ gate, which is implemented by braiding the right-most MZM from the first chain with the left-most MZM of the second chain. This is most convenient type of gate because for the spin-mapped representation, this does not involve high order spin operations to be performed.  For example, Eq. (S82) in the Supplemental Material is a second order operation, while Eq. (S60) in the Supplementary involves a product of many spin operators.

As with the standard teleportation circuit, there are primarily three steps: (i) preparation of an entangled qubit between Alice and Bob; (ii) measurement of Alice's qubits in the Bell basis; (iii) classical correction at Bob, conditioned on Alice's measurement outcome.  In our circuit, the entanglement in (i) is prepared using a logical $ \sqrt{X_1 X_2} $ gate.  The Bell measurement (ii) is performed by combining an entangling operation $ \sqrt{X_2 X_3} $ with a measurement in the local basis.  Finally, as explained in the main text, the classical correction is performed by applying two $ \sqrt{Z_3} $ gates to perform a $ Z_3 $ correction, two  $ \sqrt{X_3 X_4} $ gates with an ancilla qubit set to $ X_4 = 1$ to perform the $ X_3 $ correction.

\begin{table}
\centering
\begin{tabular}{|c|c|}
\hline
Measurement outcome    & Correction \\
\hline
$\ket{00_L}_{12} $ & $Z_3$        \\
$\ket{01_L}_{12}$ & $X_3$        \\
$\ket{10_L}_{12}$ & $X_3 Z_3$       \\
$\ket{11_L}_{12}$ & $I_3 $   \\
\hline
\end{tabular}
\caption{Classical correction required for the teleportation protocol.  \label{tab:corr} }
\end{table}

Working in the logical space, the entanglement preparation step produces the state
\begin{align}
     | E \rangle_{23} & = \sqrt{X_2 X_3} | 0 0_L \rangle_{23} \nonumber \\
     & =
     \frac{1}{\sqrt{2}} ( | 0 0_L \rangle_{23} + i | 1 1_L \rangle_{23}) .
\end{align}
Logical qubit 1 is meanwhile prepared in the state
\begin{align}
    | \psi \rangle_1  =  \alpha | 0_L \rangle_1 + \beta | 1_L \rangle_1 .
\end{align}
Now applying the $ \sqrt{X_1 X_2} $ gate we have
\begin{align}
& \sqrt{X_1 X_2}  | \psi \rangle_1   | E \rangle_{23} \nonumber \\
=  & \frac{1}{2}(\alpha \ket{000_L} + i \alpha \ket{011_L} + \beta \ket{100_L} + i \beta \ket{111_L} \nonumber \\
& +
      i\alpha \ket{110_L} - \alpha \ket{101_L} + i \beta \ket{010_L} - \beta \ket{001_L}) \\
= &\frac{1}{2}(\ket{00_L}(\alpha \ket{0_L} - \beta \ket{1_L}) + i \ket{01_L}(\beta \ket{0_L} + \alpha \ket{1_L}) \nonumber \\
& -\ket{00_L}(\beta \ket{0_L} - \alpha \ket{1_L}) + i\ket{01_L}(\alpha \ket{0_L} + \beta \ket{1_L})).
\label{telefinal}
\end{align}
A measurement in the logical basis on the first two qubits collapses the state to four outcomes, and leaves logical qubit 3
in one of four possible states, as can be seen from (\ref{telefinal}).  These can be corrected to the original state by applying the operations as summarized in Table \ref{tab:corr}.

\section{Quantum gates}

In our approach, we execute the teleportation circuit as shown in Fig.~2 of the main text by following the same steps as that followed in a topological quantum computation. All the steps of the quantum teleportation are performed by successive braiding operations and measurements.  Each of the braiding operations are performed by applying the corresponding unitary operations as derived in the previous section. Since our superconducting quantum processor is composed of spins, rather than real anyons, we perform the corresponding unitary operation that achieves the same operation to the braid.

In this section we provide the details on how these operations are translated to physical qubit operations in Fig.~2(c) of the main text. From this figure it can be seen that the only gates that are required are the $ \sqrt{X_i X_j}$, $\sqrt{Z}$, encoder, and decoder circuits.  We show that the gate decompositions as shown in Fig.~2(c) reproduce these operations. We derive these for Kitaev chains are of length $ N = 2$, according to our implementation.   We derive in this section the gates for the encoding and decoding operations which produce the states in terms of the spin-mapped MZM ground states of the Kitaev chain.  Finally, we also comment on the gates that are performed on the fourth ancilla qubit which helps to perform the $ X_3 $ classical correction.

\subsection{Logical $\sqrt{X_1 X_2}$ braiding gate}

From (\ref{braid_xx_spin_rep}) we see that the desired braiding operator acting on the physical qubits for the case $ N = 2 $ is
\begin{align}
    B_{(1,r)(2,\ell)}= \exp ( i \frac{\pi}{4} \sigma^{x}_{2}  \sigma^{x}_{3} ),  \label{sqrtxxgate}
\end{align}
The above relation was derived between chain 1 and chain 2, but more generally, the operations are applied on the right-most spin of the first chain and the left-most spin of the second chain.  Let us more generally denote $ \sigma^{\xi}_a $ as the right-most site of the first chain and $  \sigma^{\xi}_b $ as the left-most site of chain 2, where $ \xi \in \{x,y,z \}$.

On our superconducting quantum processor, the naturally available gates are CZ and single qubit unitary oeprations. Hence rather than decompose our operations into elementary CNOT gates, we perform decompositions with preference of using CZ gate instead. The CZ gate between qubits $ i $ and $ j $ can be decomposed as
\begin{align}
    \text{CZ}_{ij} = e^{i\frac{\pi}{4}}  \exp (-i\frac{\pi}{4}\sigma^z_i ) \exp(-i\frac{\pi}{4}\sigma^z_j) \exp(i \frac{\pi}{4} \sigma^z_i \sigma^z_j ) .
    \label{czdecomp}
\end{align}
By removing the single qubit gates and rotating the basis of the interaction, we can thus produce the desired braiding gate (\ref{sqrtxxgate}).  The above relation was derived for chain 1, but more generally for a chain of length $ N =2$, the operations are applied on the two spin comprising the chain.  Let us more generally denote $ \sigma^{\xi}_a $ as the left-most site and $  \sigma^{\xi}_b $ as the right-most site, where $ \xi \in \{x,y,z \}$.

The braiding gate is then
\begin{align}
 & B_{(1,r)(2,\ell)}= e^{i\frac{\pi}{4}  \sigma^x_a \sigma^x_b} \\
&= e^{i\frac{\pi}{4} \sigma^y_a} e^{i\frac{\pi}{4} \sigma^y_b } e^{i\frac{\pi}{4} \sigma^z_a \sigma^z_b} e^{-i\frac{\pi}{4} \sigma^y_a } e^{-i\frac{\pi}{4} \sigma^y_b } \\
&=e^{-i\frac{\pi}{4}} e^{i\frac{\pi}{4} \sigma^y_a} e^{i\frac{\pi}{4} \sigma^y_b } e^{i\frac{\pi}{4} \sigma^z_a } e^{i\frac{\pi}{4} \sigma^z_b} \text{CZ}_{ab}  e^{-i\frac{\pi}{4} \sigma^y_a } e^{-i\frac{\pi}{4} \sigma^y_b }  \\
&=e^{-i\frac{\pi}{4}}  R^y_b (\frac{\pi}{2}) R^y_b(\frac{\pi}{2}) R^z_a(\frac{\pi}{2}) R^z_b(\frac{\pi}{2}) \text{CZ}_{ab}  R^y_a(-\frac{\pi}{2}) R^y_b(-\frac{\pi}{2}),
\end{align}
where in the last line we have rewritten the qubit operations in terms of rotation angles on the Bloch sphere
\begin{align}
    R^\xi_j(\theta)  = \exp(i \sigma^\xi_j  \theta/2) .
    \label{rotdef}
\end{align}
where $ \xi \in \{ x,y,z \}$. The above expression gives the gate decomposition in Fig.~2(c) of the main text.

\subsection{Logical $\sqrt{Z_1}$ braiding gate}

From (\ref{physicalsqrtz1}) we see that the desired braiding operator acting on the physical qubits for $ N =2 $
\begin{align}
    B_{(1,\ell)(1,r)}= \exp ( i \frac{\pi}{4} \sigma^{y}_{1}  \sigma^{y}_{2} ).  \label{sqrtzgate}
\end{align}
The above relation was derived for chain 1, but more generally for a chain of length $ N =2$, the operations are applied on the two spins comprising the chain.  Let us more generally denote $ \sigma^{\xi}_a $ as the left-most site and $  \sigma^{\xi}_b $ as the right-most site, where $ \xi \in \{x,y,z \}$.

Analogously to the previous section, we modify (\ref{czdecomp}) into the correct form by applying single qubit gates and performing a $\sigma^x$-rotation. The braiding gate is then
\begin{align}
& B_{(1,\ell)(1,r)} =  e^{i\frac{\pi}{4} \sigma^y_a \sigma^y_b } \\
&=  e^{i\frac{\pi}{4} \sigma^x_a} e^{i\frac{\pi}{4} \sigma^x_b } e^{i\frac{\pi}{4} \sigma^z_a \sigma^z_b } e^{-i\frac{\pi}{4}\sigma^x_a } e^{-i\frac{\pi}{4} \sigma^x_b } \\
&= e^{-i\frac{\pi}{4}} e^{i\frac{\pi}{4} \sigma^x_a } e^{i\frac{\pi}{4} \sigma^x_b }  e^{i\frac{\pi}{4}\sigma^z_a } e^{i\frac{\pi}{4}\sigma^z_b } \text{CZ}_{ab}  e^{-i\frac{\pi}{4}\sigma^x_a } e^{-i\frac{\pi}{4}\sigma^x_b } \\
&= e^{-i\frac{\pi}{4}}  R^x_a(\frac{\pi}{2}) R^x_b (\frac{\pi}{2}) R^z_a (\frac{\pi}{2}) R^z_b (\frac{\pi}{2}) \text{CZ}_{ab}  R^x_a(-\frac{\pi}{2}) R^x_b (-\frac{\pi}{2}) ,
\end{align}
where in the last line we rewrote the gates in terms of (\ref{rotdef}).   The above expression gives the gate decomposition in Fig.~2(c) of the main text.

\subsection{Encoder circuit}

In this section we derive the encoder quantum circuit, defined as the unitary operation that achieves the following
\begin{align}
U_{\text{enc}} | 0 \rangle (\alpha |0 \rangle + \beta |1 \rangle ) = \alpha |0_L \rangle + \beta  |1_L \rangle ,
\label{encdef}
\end{align}
where
\begin{align}
| 0_L \rangle & = \frac{1}{\sqrt{2}} ( \ket{++} + \ket{--} ) \nonumber \\
| 1_L \rangle & = \frac{1}{\sqrt{2}} (\ket{++} - \ket{--}) .
\label{logicalmzm}
\end{align}
The encoder circuit shown in Fig.~2(c) corresponds to the operator
\begin{align}
U_{\text{enc}} &= H_2 \text{CZ}_{12}  H_1 H_2 .
\end{align}

We show explicitly this achieves (\ref{encdef}) according to the steps below
\begin{align}
U_{\text{enc}} & | 0 \rangle (\alpha |0 \rangle + \beta |1 \rangle ) = H_2 \text{CZ}_{12} (\alpha\ket{++} + \beta\ket{+-})) \nonumber  \\
= &\frac{1}{2} H_2 \text{CZ}_{12} \big[
\alpha (\ket{00} + \ket{01} + \ket{10} + \ket{11}) \nonumber \\
  &  + \beta (\ket{00} - \ket{01} + \ket{10} - \ket{11}))  \big] \\
= &\frac{1}{2} H_2 \big[
\alpha (\ket{00} + \ket{01} + \ket{10} - \ket{11}) \nonumber \\
  &  + \beta (\ket{00} - \ket{01} + \ket{10} + \ket{11}))  \big] \\
  = &\frac{1}{\sqrt{2}} \big[
\alpha (\ket{00}  + \ket{11})   + \beta (\ket{01} + \ket{10}))  \big] \\
= &\frac{1}{\sqrt{2}}(\alpha(\ket{++} + \ket{--}) + \beta(\ket{++} - \ket{--}))) \\
= & \alpha\ket{0_L} + \beta\ket{1_L} , \label{finalencstate}
\end{align}
as desired.

\subsection{Decoder circuit and error detection}

Similarly, we also need to be able to perform reverse operation, where the input state is the a two qubit MZM encoded state $ \alpha\ket{0_L} + \beta\ket{1_L})$, and output the unencoded qubit state. This is of course the inverse of  (\ref{encdef}) and given by
\begin{align}
U_{\text{dec}} &= U_{\text{enc}}^\dagger \nonumber \\
& = H_1^\dagger  H_2^\dagger \text{CZ}_{12}^\dagger H_2^\dagger \nonumber \\
& = H_1 H_2 \text{CZ}_{12} H_2 ,
\end{align}
since the Hadamard and CZ operations are Hermitian.

As discussed in the main text, when a phase flip occurs on the logical states (\ref{logicalmzm}), the states transform as
\begin{align}
 |\tilde{0}_L \rangle & = \sigma^z_1 | 0_L \rangle =  \sigma^z_2  | 0_L \rangle =
\frac{1}{\sqrt{2}} ( \ket{-+} + \ket{+-}) \nonumber \\
 |\tilde{1}_L \rangle & = \sigma^z_1  | 1_L \rangle =  -\sigma^z_2 | 1_L \rangle =
\frac{1}{\sqrt{2}} ( \ket{-+} - \ket{+-} ) .
\end{align}
We now show that decoding a state with a single phase flip error results in a $ |1 \rangle $ on the first qubit, which allows one to detect the error.

Specifically, we consider that a $ \sigma^z_1 $ error occurs on the output state (\ref{finalencstate}) such that we have the state
\begin{align}
\sigma^z_1 (\alpha | 0_L \rangle + \beta | 1_L \rangle ) & = \alpha | \tilde{0}_L \rangle + \beta | \tilde{1}_L \rangle  \nonumber \\
& = \frac{1}{\sqrt{2}} [ ( \alpha + \beta)  \ket{-+}
+ ( \alpha - \beta)  \ket{+-} ]
\end{align}
Applying the decoder operation then gives
\begin{align}
U_\text{dec} &  (  \alpha | \tilde{0}_L \rangle + \beta | \tilde{1}_L \rangle)  \nonumber \\
= & \frac{1}{\sqrt{2}} H_1 H_2 \text{CZ}_{12}  [ ( \alpha + \beta)  \ket{-0}  + ( \alpha - \beta)  \ket{+1} ] \nonumber \\
= & \frac{1}{2}  H_1 H_2 \text{CZ}_{12}  [  ( \alpha + \beta) ( \ket{00} - \ket{10}) + ( \alpha - \beta)( \ket{01} + \ket{11} ] \nonumber \\
= & \frac{1}{2}  H_1 H_2 [  ( \alpha + \beta) ( \ket{00} - \ket{10}) + ( \alpha - \beta)( \ket{01} - \ket{11} ] \nonumber \\
= &  H_1 H_2 (\alpha  \ket{-+}
+ \beta  \ket{--} ) \nonumber \\
= &| 1 \rangle ( \alpha  \ket{0} + \beta  \ket{1}  ) .
\end{align}
We this see that the decoder the errored state produces a state $ | 1 \rangle $ on the first qubit as claimed.  A phase flip on the second qubit gives similar results, except that $ \beta \rightarrow - \beta $.

\subsection{Ancilla qubit}

We finally comment on the gate operations performed on the fourth ancilla qubit.  As explained in the main text, the only role of this is to facilitate the $ X_3 $ classical correlation required in the teleportation circuit.  Since the braiding operations of Fig.~\ref{fig2new} does not provide a single qubit $ X $ gate, we can perform this instead by preparing a fourth ancilla qubit in the state with eigenvalue $ X_4 = +1$.  Then applying the logical $ \sqrt{X_3 X_4} $ gate twice, we accomplish the $ X_3 $ gate.

The state with eigenvalue $ X_4 = +1$ is in terms of physical qubits
\begin{align}
    |+_L \rangle & = \frac{1}{\sqrt{2}} ( | 0_L \rangle +  | 1_L \rangle ) \nonumber \\
    & = \ket{++}, \label{fourthqubit}
\end{align}
according to (\ref{logicalmzm}).  This could be prepared using the encoder of the previous section, but a simpler way is simply to apply two Hadamard gates
\begin{align}
     |+_L \rangle = H_1 H_2 \ket{00}.
\end{align}

The only operation that is applied to logical ancilla qubit 4 is the braiding operation  $ \sqrt{X_3 X_4} $, which (\ref{fourthqubit}) is an eigenstate of.  Hence it should remain unchanged after each braiding operation.

Finally, the state is decoded using $ U_\text{dec}$.  We use the decoding operation here because we would like to detect any phase flip errors that may have inadvertently occurred on these qubits.  Without any errors, the state after the decoding is
\begin{align}
   U_\text{dec} |+_L \rangle =  |0 \rangle | + \rangle
\end{align}
according to (\ref{finalencstate}).  A measurement of the second qubit here in the $ \sigma^z $ eigenbasis gives $ | 0 \rangle $ and $ | 1 \rangle $ with 0.5 probability each. Rather than obtaining a random result, it is more informative to measure in a basis such that any deviations from the ideal case can easily detected.  For this reason we use the modified decoder corresponding to
\begin{align}
    U_\text{dec}' = H_1 \text{CZ}_{12} H_2
\end{align}
such that instead the final state is
\begin{align}
   U_\text{dec}' |+_L \rangle =  |0 \rangle | 0 \rangle .
\end{align}
In this way the error detection can be still performed in a consistent way, and deviations from the ideal result of  $ | 0 \rangle $ on the second qubit can be easily detected.

\section{Simulation results of Majorana teleportation}

To numerically test our teleportation circuit we simulated the gate evolutions as given in Fig.~2 of the main text, including gate errors and dephasing effects. We model both errors by applying random gates that simulate the effect of the noise.  In order to match the experimental results we begin by tuning our numerical parameters to fixed values provided by characterization of the experiment.

To simulate the gate errors, we assume that the Hamiltonians that implement the gate are performed correctly, but there is some randomness in the time of the pulse. The time that the pulse is applied is drawn from a Gaussian distribution, and the fidelity of the simulation is calculated for each pulse duration according to
\begin{align}
   f_{\frac{X}{2}\text{S}} &= \frac{1}{N} \sum_n^N \lvert \bra{1} R_x(\pi + \xi_x \pi) \ket{0} \rvert^2 \\
   f_\text{CZS} &= \frac{1}{N} \sum_n^N \lvert \bra{1-} R_z(\xi_\text{CZ} \pi) \text{CZ}_{12} \ket{1+} \rvert^2
\end{align}
for the $ X/2 $ and $\text{CZ}_{12}$ gates respectively.  Here $ R_{x,z} $ are single qubit rotation operators, and for the $\text{CZ}_{12}$ gate the random phase is applied on the target qubit.  The gate times $ \xi $ are chosen from a Gaussian distribution with mean zero and variance $ \sigma^2$, i.e. $  \xi_x \sim \mathcal{N}(0,\,\sigma_g^{2}) $ and  $\xi_\text{CZ} \sim \mathcal{N}(0,\,\sigma_\text{CZ}^{2}) $. The parameters to tune are the standard deviations for random sampling:  $\sigma_\text{CZ}^{2}$ for a two qubit gate and $\sigma_g^{2}$ for a single qubit gate. The tuned values for each qubit are provided in the Table \ref{tab:numerical}.

Dephasing is also simulated in the same way by introducing a set of random Gaussian pulses in the middle of the processing. Again, as in case of gate error, dephasing error is characterized by a variance parameter.  We denote the variance of the randomly applied dephasing as $\sigma_d^{2}$. Appropriate values for the variance $\sigma_d^{2}$ are calculated from the experimental dephasing times $T^*_2$ as given in Table \ref{tab:QPU}. In order to adapt those experimentally obtained quantities to act in the numerical simulation we convert them to dimensionless units by applying normalization and multiplying by a common phenomenological constant $c_d$ which accounts for the overall amount of decoherence in the system and is shared among all the qubits to preserve the individual proportions resulting from experimentally measured $T^*_2$. The value of $c_d$ is calibrated to match the final simulated teleportation fidelities to experimentally obtained corresponding values.

The teleportation fidelity is calculated as follows. The initial state is a state to be teleported initialized on qubit $Q2$,
\begin{align}
    \ket{\Psi_0} &=
    \ket{0}  \otimes \ket{\psi}   \otimes \ket{0}  \otimes  \ket{0}  \otimes  \ket{0}   \otimes \ket{0}   \otimes \ket{0}  \otimes   \ket{0}
\end{align}
where $ \ket{\psi} \in \{\ket{0}, \ket{1}, \ket{+}, \ket{-}, \ket{+i}, \ket{-i}\}$ is the state to be teleported. From this initial state, we calculate the fidelity by applying the unitary teleportation circuit $U_n(\sigma_d, \sigma_g, \sigma_\text{CZ})$ with random gate errors and random dephasing with standard deviations $\sigma_g$, $\sigma_\text{CZ}$ and $\sigma_d$. The index $ n $ represents the $n$th random draw. To the resulting state we apply a classical correction circuit $U^c_n(\sigma_g, \sigma_\text{CZ})$. The calculation is repeated for all possible classical corrections and all possible measurements of error detecting qubits. This is performed by applying a series of projectors, which gives the final state of the form
\begin{align}
    \ket{\psi^f_{c,m,k,n}} &= \Pi_{m,k} U^c_n(\sigma_g, \sigma_\text{CZ}) \Pi_c U_n(\sigma_d, \sigma_g, \sigma_\text{CZ}) \ket{\Psi_0}
    \label{psitilde}
\end{align}
where the projectors are
\begin{align}
    \Pi_c = &I \otimes \dyad{c_1} \otimes I \otimes \dyad{c_2} \nonumber \\
    \otimes &I \otimes I \otimes I \otimes I \\
    \Pi_{m,k} = &\dyad{m_1} \otimes I \otimes \dyad{m_2} \otimes I \nonumber \\
    \otimes &\dyad{m_3} \otimes I \otimes \dyad{m_4} \otimes \dyad{k}
\end{align}
Here the index $ c $ runs over all classical correction outcomes, and $ m $ runs over the the measurements over the syndrome measurements, $k$ representing the measurement of the ancilla qubit, which plays no role in the computation thus no post-selection is defined on its measured value.  We note that the state (\ref{psitilde}) is unnormalized due to the projectors acting on it.

We now explain how the teleportation fidelities are calculated from the state including the gate and dephasing errors.  First consider the case when no error syndrome measurements are made (NS).  Given all possible classical corrections, all possible error detecting qubit outcomes and all possible ancilla qubit outcomes, that have been evaluated, for $n$th random draw we can prepare for a traced out density matrix corresponding to the teleported qubit.
\begin{align}
    \rho_{6,n}^{\text{NS}} &= \Tr_{1,2,3,4,5,7,8} (\sum_{c}\sum_{k}\sum_{m}\dyad{\psi^f_{c,m,k,n}}) \label{trace_out}
\end{align}
For the case that error syndrome measurements are made (ES), we fix the outcomes of the odd numbered qubits to outcome zero $ m_1 = m_2 = m_3 = m_4 = 0$
\begin{align}
    \rho_{6,n}^{\text{ES}} &= \Tr_{1,2,3,4,5,7,8} (\sum_{c}\sum_{k} \dyad{\psi^f_{c,m=0,k,n}}) . \label{trace_out2}
\end{align}
We note that the above is an unnormalized state because the full set of measurements are not used.

By averaging over a large number of random draws to simulate the effects of gate errors and decoherence, and applying appropriate normalization we get the fidelity of the teleported state $\ket{\psi}$
\begin{align}
    f_S &= \frac{1}{N} \sum_{n=1}^N \frac{\expval{\rho_{6,n}}{\psi}}{\tr(\rho_{6,n})} .
\end{align}
The denominator is present to account for the case that the state is unnormalized.

We calculated the fidelity for both cases, with and without error detection, for all input state $\ket{\psi}$. The numerical values we obtained compared against experimental values after averaging over $N = 2000$ random runs are provided in the Table \ref{tab:fidelity}. The overall features of fidelity profile matches the experiment and in average among all the input states, the error detected fidelity is above the $\frac{2}{3}$ threshold. We observed the closest match of the fidelities for the constant $c_d=0.15$. Generally the theoretically calculated fidelities are higher than the experimentally obtained values. We attribute this to the fact that measurement errors are not taken into account in our simulation.  We expect that this will further reduce the overall fidelities.

\begin{figure}
\begin{center}
\includegraphics[width=\linewidth]{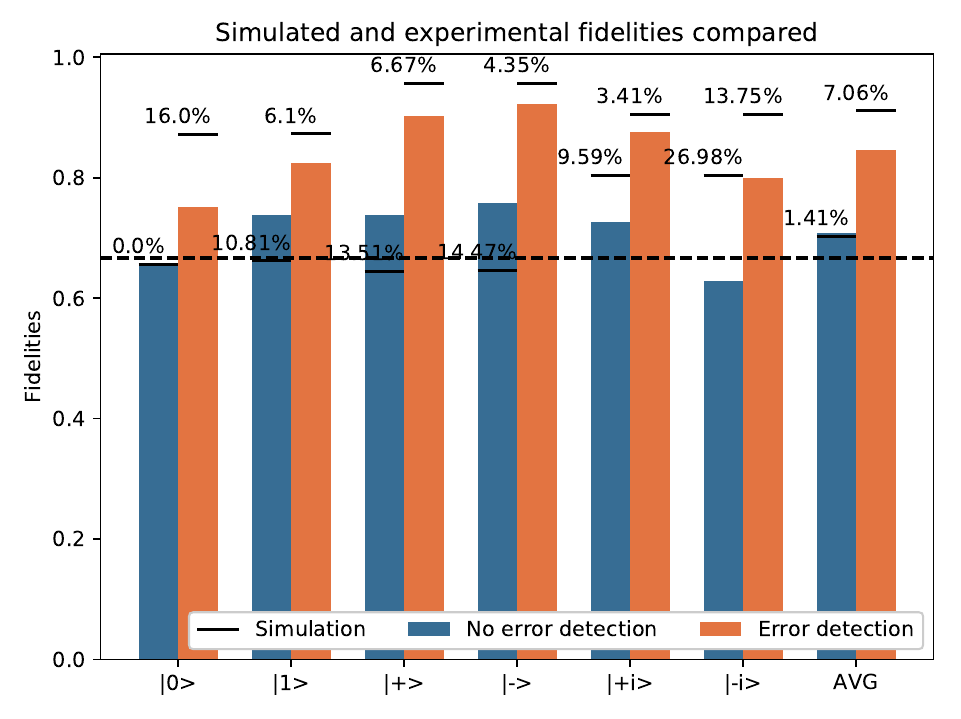}
\end{center}
\caption{Bar chart visualization of experimental and numerical fidelity from Table \ref{tab:fidelity}, for cases with and without error detection. Chart includes percent errors to compare how closely simulation matches the experiment. The horizontal dashed line is indicating the $\frac{2}{3}$ threshold. \label{fignumerics}}
\end{figure}

The detailed values of the simulated and experimental fidelities, as well as the errors and the averages are provided in the Table \ref{tab:fidelity}.  The same data is visualized in form of a bar chart on Fig.  \ref{fignumerics}.

\begin{table*}[htbp]
\centering
\begin{tabular}{l c c}
\hline \hline
Description & Experiment & Simulation \\ \hline
Teleportation without error detection & $f_\text{NE}$ & $f_\text{NS}$ \\
Teleportation with error detection & $f_\text{EE}$ & $f_\text{ES}$ \\ \hline
X/2 gate fidelity & $f_{\frac{X}{2}\text{E}}$ & $f_{\frac{X}{2}\text{S}}$  \\
CZ gate fidelity & $f_\text{CZE}$ & $f_\text{CZS}$  \\
\hline \hline
\end{tabular}
 \caption{\small{\textbf{Fidelity notation} to assign a dedicated symbol to a fidelity value corresponding to particular scenario.}}
   \label{tab:fidelity_notation}
\end{table*}

\begin{table*}[htbp]
\centering
\begin{tabular}{l l l l l l l l}
\hline \hline
& $\ket{0}$ & $\ket{1}$ & $\ket{+}$ & $\ket{-}$ & $\ket{+i}$ & $\ket{-i}$ & AVG \\ \hline
$f_\text{NE}$ &0.66 & 0.74 & 0.74 & 0.76 & 0.73 & 0.63 & 0.71 \\
$f_\text{NS}$& 0.66 & 0.66 & 0.64 & 0.65 & 0.8 & 0.8 & 0.7 \\
Error (\%) & 0.0 & 10.81 & 13.51 & 14.47 & 9.59 & 26.98 & 1.41 \\ \hline
$f_\text{EE}$ &0.75 & 0.82 & 0.9 & 0.92 & 0.88 & 0.8 & 0.85 \\
$f_\text{ES}$ & 0.87 & 0.87 & 0.96 & 0.96 & 0.91 & 0.91 & 0.91 \\
Error (\%) & 16.0 & 6.1 & 6.67 & 4.35 & 3.41 & 13.75 & 7.06 \\
\hline \hline
\end{tabular}
 \caption{\small{\textbf{Teleportation fidelity} for a set of input states and average between all those states, calculated with and without error detection, compared to the experimental fidelity for same input states by calculating the percent error.}}
   \label{tab:fidelity}
\end{table*}

\begin{table*}[htbp]
\centering
\begin{tabular}{l l l l l l l l l l}
\hline \hline
Qubit & Q1 & Q2 & Q3 & Q4 & Q5 & Q6& Q7 & Q8 & AVG \\ \hline
$\sigma_d^2$ &0.11407 &0.05426 &0.11841 &0.03014 &0.15    &0.05764 &0.11334 &0.06969 &0.08844  \\
$\sigma_g^2$ &0.016 & 0.017 & 0.017 & 0.018 & 0.014 & 0.017 & 0.013 & 0.014 & 0.01575 \\
$f_{\frac{X}{2}\text{S}}$ & 0.9994 & 0.9993 & 0.9993 & 0.9992 & 0.9995 & 0.9993 & 0.9996 & 0.9995 & 0.9994 \\
$f_{\frac{X}{2}\text{E}}$ & 0.9994 & 0.9993 & 0.9993 & 0.9992 & 0.9995 & 0.9993 & 0.9996 & 0.9995 & 0.9994 \\
\hline \hline
$\sigma_\text{CZ}^2$ & \multicolumn{9}{l}{~~~~0.08287   0.075524  0.0729  ~0.0757    0.10285   0.0528  ~~~0.056    ~~~~~~~0.074092} \\
$f_\text{CZS}$ & \multicolumn{9}{l}{~~~~~0.9832   ~0.9861  ~~0.987  ~~0.9861    ~~0.9744   ~~0.9932  ~~~0.9923    ~~~~~0.986} \\
$f_\text{CZE}$ & \multicolumn{9}{l}{~~~~~0.983   ~~~0.986  ~~~0.987  ~~0.986    ~~~~0.974   ~~~0.993  ~~~~0.992    ~~~~~~ 0.986} \\

\hline \hline
\end{tabular}
 \caption{\small{\textbf{Numerical calibration} of qubits to match the experimental performance. Includes gate fidelity of of each qubit, the standard deviation of random error used to reproduce the effect of dephasing.}}
   \label{tab:numerical}
\end{table*}

\section{Experimental details}

Our superconducting quantum processor has 12 frequency-tunable transmon qubits of the Xmon variety.  In our experiments, the eight qubits (Fig.~1(a)) are chosen from the quantum processor. The processor has qubits lying on a 1D chain, and the qubits are capacitively coupled to their nearest neighbors. Each qubit has a microwave drive line (XY), a fast flux-bias line (Z) and a readout resonator. All readout resonators are coupled to a common transmission line for state readout. The single-qubit rotation gates are implemented by driving the XY control lines, and the CZ gate is implemented by driving the Z line using the ``fast  adiabatic'' method. The performances of the eight qubits we chosen in our experiment are listed in Table \ref{tab:QPU}.

 \begin{table*}[htbp]
\centering
\begin{tabular}{l l l l l l l l l l}
\hline \hline
Qubit & Q1 & Q2 & Q3 & Q4 & Q5 & Q6& Q7 & Q8 & AVG \\ \hline
$\omega_{10}/2\pi$ (GHz) &5.066& 4.18 & 5.01 & 4.134 & 5.08 & 4.22 & 5.132 &4.19 &\multicolumn{1}{c}{-} \\
$T_1$ ($\mu s$) &35.2 & 31.69 & 35.23 & 31.01 & 25.79 & 27.98 & 34.79 & 28.94 & 31.32 \\
$T_2^*$ ($\mu s$) & 4.73 & 2.25 & 4.91 & 1.25 & 6.22 & 2.39 & 4.7 &2.89 & 3.67 \\
$f_{00}$ &0.980 & 0.952 & 0.981 & 0.949 & 0.923 & 0.896 & 0.915 &0.912 & 0.939 \\
$f_{11}$ &0.865 & 0.866 & 0.905 & 0.887 & 0.863 & 0.858 & 0.888 &0.873 & 0.876 \\ \hline
X/2 gate fidelity &0.9994 & 0.9993 & 0.9993 & 0.9992 & 0.9995 & 0.9993 & 0.9996& 0.9995& 0.9994 \\
CZ gate fidelity & \multicolumn{9}{l}{~~~~~0.983   ~~0.986  ~~0.987  ~~0.986    ~~0.974   ~~0.993  ~~0.992    ~~~~ 0.986} \\ \hline \hline
\end{tabular}
 \caption{\small{\textbf{Performance of qubits}. $\omega_{10}$ is idle points of qubits. $T_1$ and $T_2^*$ are the energy relaxation time and dephasing time, respectively. $f_{00}$ ($f_{11}$) is the possibility of correctly readout of qubit state in $\ket{0}$ ($\ket{1}$) after successfully initialized in $\ket{0}$ ($\ket{1}$) state. X/2 gate fidelity and CZ gate fidelity are single and two-qubit gate fidelities obtained via performing randomized benchmarking.       }}
   \label{tab:QPU}
\end{table*}

\begin{figure*}
\begin{center}
\includegraphics[width=\linewidth]{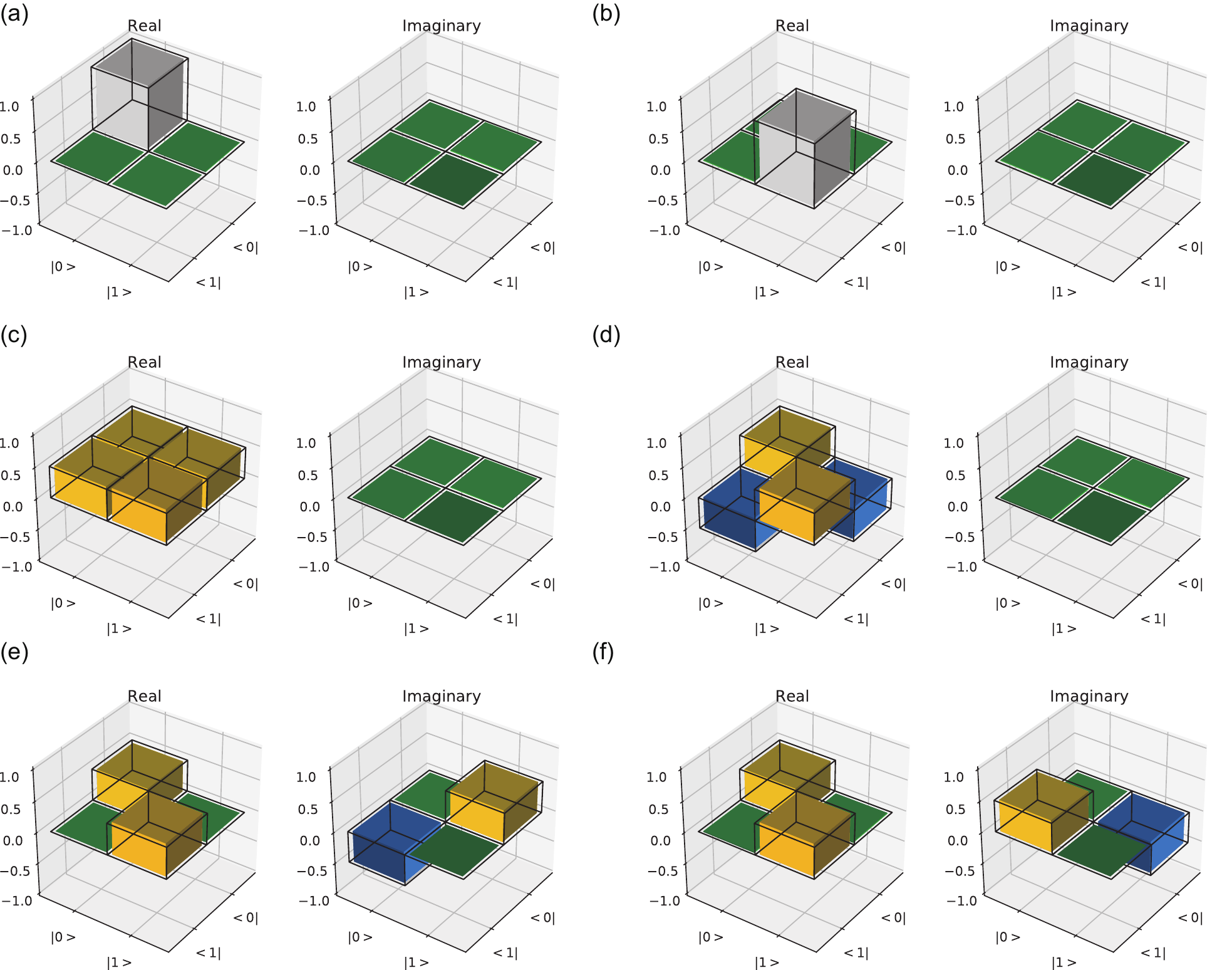}
\end{center}
\caption{Tomography of the initial state ${\rm{|}}\psi {\rangle _2}$. The initial state prepared on qubit 2 is (a) $ | 0 \rangle $,  (b) $ | 1 \rangle $,  (c) $ | + \rangle $,  (d) $ | - \rangle $,  (e) $ | +i \rangle $,  (f) $ | -i \rangle $.
\label{figtomography_init}}
\end{figure*}

During running the quantum circuits, we have performed the tomography measurement on the initial state ${\rm{|}}\psi {\rangle _2}$ on qubit 2 that we prepared for teleportation (see Fig.~\ref{figtomography_init}), and the fidelities of six initial states are 0.9998, 0.9998, 0.9982, 0.9997, 0.9999, and 0.9989.

\begin{figure*}
\begin{center}
\includegraphics[width=\linewidth]{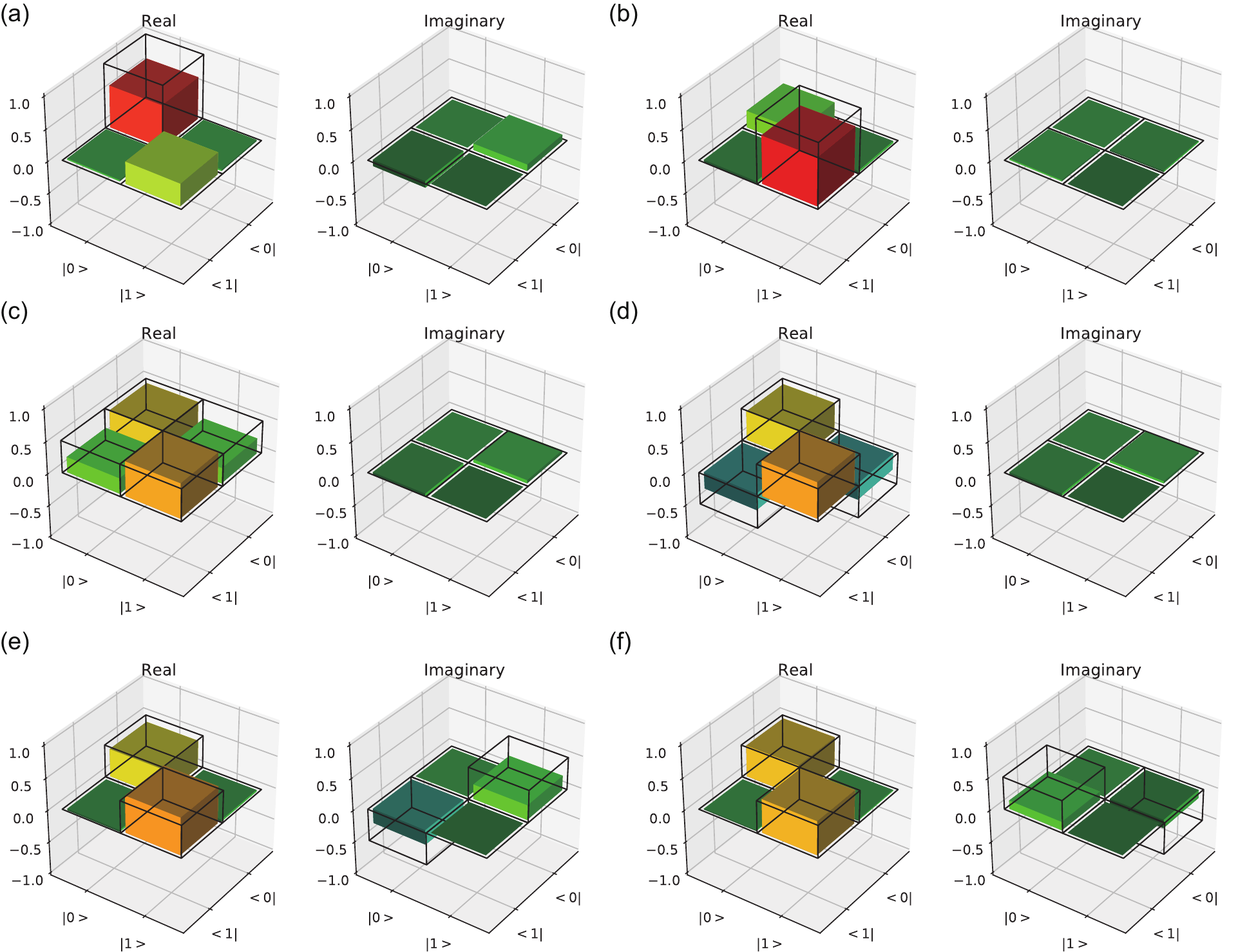}
\end{center}
\caption{Tomography of the final teleported state before using the error syndrome measurements. The initial state prepared on qubit 2 is (a) $ | 0 \rangle $,  (b) $ | 1 \rangle $,  (c) $ | + \rangle $,  (d) $ | - \rangle $,  (e) $ | +i \rangle $,  (f) $ | -i \rangle $. Frames show ideal teleportation states, colored bars shows the experimentally determined state.
\label{figtomography_no_detect}}
\end{figure*}

In addition, we also performed the tomography measurement on the final teleported state that before using the error syndrome measurements (see Fig.~\ref{figtomography_no_detect}).

\bibliographystyle{apsrev}
\bibliography{paperrefs}